\documentclass[preprint,5p,10pt]{elsarticle}

\usepackage{amsmath,amssymb}
\usepackage{booktabs}
\usepackage{url}
\usepackage{geometry}
\usepackage[colorlinks,bookmarksopen,bookmarksnumbered,citecolor=blue, linkcolor=blue, urlcolor=blue]{hyperref}

\journal{Physics of Dark Universe}

\usepackage{soul,bm}

\begin{document}

\begin{frontmatter}


 \title{Toward a direct measurement of the cosmic acceleration: the first preparation with FAST}
 
\author[label1,label2]{Chang-Zhi Lu}
\author[label1,label2]{Kang Jiao}
\author[label3]{Tingting Zhang\corref{cor1}}
\ead{101101964@seu.edu.cn}
\author[label1,label2,label4]{Tong-Jie Zhang\corref{cor1}}
\cortext[cor1]{Corresponding author}
\ead{tjzhang@bnu.edu.cn}
\author[label5]{Ming Zhu}

\affiliation[label1]{organization={Institute for Frontiers in Astronomy and Astrophysics, Beijing Normal University},
	postcode={102206},
	state={Beijing},
	country={China}}
\affiliation[label2]{organization={Department of Astronomy, Beijing Normal University},
	postcode={100875},
	state={Beijing},
	country={China}}
\affiliation[label3]{organization={College of Command and Control Engineering, PLA Army Engineering University},
	postcode={210000}, 
	state={Nanjing},
	country={China}}
\affiliation[label4]{{Institute for Astronomical Science, Dezhou University},
	postcode={253023}, 
	state={Dezhou},
	country={China}}
\affiliation[label5]{organization={National Astronomical Observatories, Chinese Academy of Sciences},
	postcode={100101}, 
	state={Beijing},
	country={China}}




\begin{abstract}
Damped Lyman-$\alpha$ Absorber(DLA) of HI 21cm system is an ideal probe to directly measure cosmic acceleration in real-time cosmology via Sandage-Loeb(SL) test. During short observations toward two DLAs in the commissioning progress of FAST, we manage to exhibit an HI 21cm absorption feature from PKS1413+135 spectrum in one epoch with our highest resolution up to 100 Hz, preliminarily validating the frequency consistency under different resolutions and bandwidths. We make a Gaussian fitting to extract the spectral features, introduce two theoretical indicators to describe the fitted velocity uncertainty, and ultimately give a mean redshift and its constraint of $z_\mathrm{M}=0.24670045\pm0.00000036$ in accord with most literature. But our redshift error of the target is still three magnitudes higher than the level we can reach the drift signal. Though our first preparation has some flaws in time recording and diode settings, it still proves the correctness of our data process. Confined by limited observing time, we do not strech FAST's ability to obtain a better velocity constraint, so further researchs are needed and in schedule. With fine sensitivity and improving spectral resolution, such observations in FAST could have reasonable possibility to explore cosmic acceleration in late time universe practically.
\end{abstract}

%


\end{frontmatter}


\section{Introduction}
Since the outstanding supernovae research\cite{01-AJ98-Riess,02-ApJ99-Perlmutter}, plenty of theories have been conceived to explain the accelerating expansion of the current universe, which is believed to originate in dark energy. Confused by diverse possible forms of Equation of State(EoS) for dark energy, the profound mystery of expansion remains ambiguous. Nowadays our cosmological probes, SNIa\cite{03-ApJ18-Scolnic}, CMB\cite{04-AA20-Planck}, BAO\cite{05-ApJ16-HONG}, and Gravitational Wave\cite{06-Nature18-Chen}, which keep tracking the components of the universe, have made indirect observations for the qualitative cosmic acceleration in model-dependent ways.

The essence for quantitative cosmic acceleration suggests measuring the velocity change of the celestials that faithfully trace Hubble flow at different epochs and naturally have well-recognized stillness in comoving space. Sandage\cite{07-ApJ62-Sandage} proposed the first approach measuring the redshift drift of a radio galaxy. Then the technique was reformed to measure the redshift difference of a Lyman-$\alpha$ forest by Loeb\cite{08-ApJ98-Loeb}. In the FLRW metric, the SL effect is a globally dynamical probe independent of cosmic geometry and a model-independent indicator of cosmic acceleration.

Although SL effect is beyond our observational ability now,its possible applications are well-conceived by many researchers. The possible parameter space of dark energy models is so vast that the predictions of some models are easily tuned to observational results\cite{09-PASA20-Weltman}. In this case, an adequately high-precision SL-observation is powerful to distinguish from simple Rh cosmology\cite{10-MNRAS16-Melia} to complex interacting dark energy models\cite{11-PRD14-Cala,12-EPJC14-ZHANG}. Besides, it could constrain modified gravity theories\cite{13-PRD13-LI,14-NewA20-Bhatta}, provide the theoretical value of acceleration and jerk\cite{15-PHD16-Martins}, restrict curvature parameter and inflationary model\cite{16-JCAP18-Jimenez}, and partly break the degeneracies of cosmological parameters\cite{17-JCAP15-YUAN}. It was also used to study the backreaction conjecture from the inhomogeneity\cite{53-MNRAS20-Koks,54-PRL21-Koks}, research the violation of strong energy condition\cite{19-PRD21-Heinesen}, and explore the effect of redshift drift in strong gravitational lensings\cite{20-PRD17-Piattella}. In the radio Damped Lyman-$\alpha$ Absorber(DLA) path to cosmic acceleration, there have been positive developments over the years. Martins et al.\cite{55-arXiv21-Martins} provided an illuminating discussion about the sensitivity of cosmological parameters to the dimensionless velocity drift and compared the several cosmic acceleration project in Lyman-$\alpha$ forest realm via MCMC and Fisher matrix.

In the theory, based on FLRW space-time, Martins et al.\cite{15-PHD16-Martins} studied the second derivatives of redshift. Considering the pure symmetry of motion without dynamics, Lobo et al.\cite{21-JCAP20-Lobo} made a Taylor expansion of the time derivative of redshift to n-th order. Heinesen\cite{22-PRD21-Heinesen} expressed the redshift drift in the form of physical multipole moments and discussed its effect in large scale conditions.

In the simulation, Corasaniti et al.\cite{52-PRD07-Corasannti} studied the feasibility of redshift drife measurements and firstly named it as SL test. The CODEX spectrograph team of ELT\cite{25-MNRAS08-Liske} carefully considered the Lyman-$\alpha$ Forests(LAF) in high-redshift quasars($z:2\sim 5$) as an excellent accelerometer, and studied the sensitivity(accuracy) of measured radial velocity versus their redshifts and observed SNR, establishing a strict and well-accepted analysing framework in LAF approach of redshift drift, and making LAF as a mainstream in this field with solid theoretical foundation. Kloeckner et al.\cite{27-AASKA14-Kloeck} researched the cosmic acceleration observability of SKA and related systematics. Yu et al.\cite{28-PRL14-YU} forecasted that a CHIME-like survey could detect the acceleration of $\Lambda$CDM cosmology within five sigma confidence, and also researched the DLA survey capacity of Chinese new generation of telescopes: Tianlai and FAST\cite{29-RAA17-YU}. Bolejko et al.\cite{23-arXiv19-Bolejko} listed theoretical dynamical contaminations of redshift drift, suggested a method observing the drift of the subtraction of two redshifts within a negligible angular distance to reduce systematical errors, and combined redshift drift with the flux drift of the same origin to improve the availability of observation data. Recently there are many LAF-approach experiments proposed apart from CODEX team of ELT, such as the space-borne and low-cost Cosmic Accelerometer experiment\cite{59-BAAS19-Eiken}, the Accelerometer Programme\cite{24-MNRAS19-Cooke} which aims to measure the redshift drift between two different non-zero redshifts instead of comparing to today via ELT. Besides, the current generation spectrographs like ESPRESSO and NEID are expected to achieve the radial valocity precision of $\sim$ 10 cm/s\cite{60-arXiv22-Chakra}, which would greatly contribute to redshift drift observations through LAF approach.

In the observation of radio DLA approach, Darling\cite{26-ApJ12-Darling} observed more than ten DLAs over a decade and obtained the best constraint of redshift drift so far. Although the constraint was three orders of magnitude larger than the theoretical prediction, it confirmed the long-term frequency stability of these DLAs and GBT telescope, indicating that expanding the samples and prolonging the time baseline can reach higher precision. Jiao et al.\cite{30-JCAP20-JIAO} distinguished three related concepts in cosmic acceleration, modified the expectation of the FAST DLA survey, proposed a combined observation mode for DLA, and made a comparative observation as a technical pathfinder of cosmic acceleration.

Moreover, another similar idea to redshift drift, the differential age of cosmic chronometers proposed by Jimenez and Loeb\cite{61-apj02-Jimen}, measures cosmic expansion at higher redshifts from the relative ages of massive, passively evolving and old galaxies which formed at a roughly same time and are separated by a small redshift interval $\Delta z$, inferring the derivative $dz/dt$ from the ratio of $\Delta z$ to the age difference $\Delta t$. This derivative could help to constrain the cosmological model parameters and even the Hubble parameter, e.g. \cite{17-JCAP15-YUAN,62-MN22-RuizZ,63-arx22-Jiao}.

Aiming to extract the precise location of HI 21cm absorption line, and to examine the detectability of the line drift in the future, we introduce two indicators in Gaussian fitting to express the fitted velocity uncertainty, and discuss some critical challenges in observing cosmic acceleration(SL effect) through HI 21cm absorption systems. In section \ref{sec2}, we briefly talk about the theories about the acceleration, DLA, systematics and indicators we use. We explain our observational settings on FAST and data process in section \ref{sec3.1}, \ref{sec3.2} respectively, give results in section \ref{sec3.3} and make discussions in section \ref{sec3.4}. We place our conclusions in section \ref{sec4}.

\section{Methodology}\label{sec2}
\subsection{Cosmic acceleration signal} 
The related formulae derivation of cosmic acceleration were well-described. Here we revise some of their results briefly. 

In a $\Lambda$CDM cosmology with a cosmological constant, the expansion rate of the universe is defined as:
\begin{equation}
	E(z)=\sqrt{\Omega_{\mathrm{r}0}(1+z)^4+\Omega_{\mathrm{m}0}(1+z)^3+\Omega_{\mathrm{k}0}(1+z)^2+\Omega_{\mathrm{\Lambda}0}},
\end{equation}
where $\Omega_{\mathrm{r}0},\Omega_{\mathrm{m}0},\ \Omega_{\mathrm{k}0},\ \Omega_{\mathrm{\Lambda}0}$ represent the today density parameters of radiation, matter, curvature and dark energy(in a constant form) separately.

Making a Taylor expansion\cite{08-ApJ98-Loeb} of the scale factor $a(t+\Delta t)\approx a(t)+\dot{a}(t)\Delta t$ in the redshift drift $\Delta z \equiv a(t_\mathrm{o}+\Delta t_\mathrm{o})/a(t_\mathrm{e}+\Delta t_\mathrm{e}) - a(t_\mathrm{o})/a(t_\mathrm{e})$, or taking the total derivative of redshift with respect to observation time\cite{25-MNRAS08-Liske,15-PHD16-Martins}, we can derive the formula of redshift drift $\Delta z$:

\begin{equation}
	\frac{\Delta z}{\Delta t_\mathrm{o}}=(1+z)H_0-H(z)=[1+z-E(z)]H_0.
\end{equation}

Conventionally we transform the redshift drift to a spectroscopical velocity drift\cite{32-MNRAS19-Alves}:
\begin{equation}
	\Delta v=\frac{c\Delta z}{1+z}=c[1-\frac{E(z)}{1+z}]H_0\Delta t_o.
\end{equation}

Measuring cosmic acceleration from raw spectra requires very high spectral resolution. Two epochs observations toward a same DLA will give $\nu_{\mathrm{ob}1}=\nu_\mathrm{em}/(1+z),\nu_{\mathrm{ob}2}=\nu_\mathrm{em}/(1+z+\Delta z)$, where $\nu_\mathrm{em}=1420.405751768\ \text{MHz}$ is the frequency of HI 21cm radiation in the rest frame, $z$ is the first measured redshift, $\Delta z$ is the redshift drift. Suppose a ten-year accumulation $\Delta z_{10}$, the frequency drift is:
\begin{equation} 
	\Delta\nu_{10}(z)=\nu_{\mathrm{ob}2}-\nu_{\mathrm{ob}1}\approx\nu_\mathrm{em}\frac{\Delta z_{10}}{(1+z)^2}.
\end{equation}
For the line width in the rest frame $\Delta v_\mathrm{rst}$ is much less than the speed of light, we can turn the frequency drift to the radial velocity drift according to Meyer et al. \cite{43-PASA17-Meyer}:
\begin{equation} 
	\Delta v_{10}(z)=(1+z)\Delta v_\mathrm{rst}\approx\frac{c(1+z)^2}{\nu_\mathrm{em}}\Delta\nu_{10}(z).
\end{equation}

\begin{figure}[htb]
	\centering
	\includegraphics[scale=0.6]{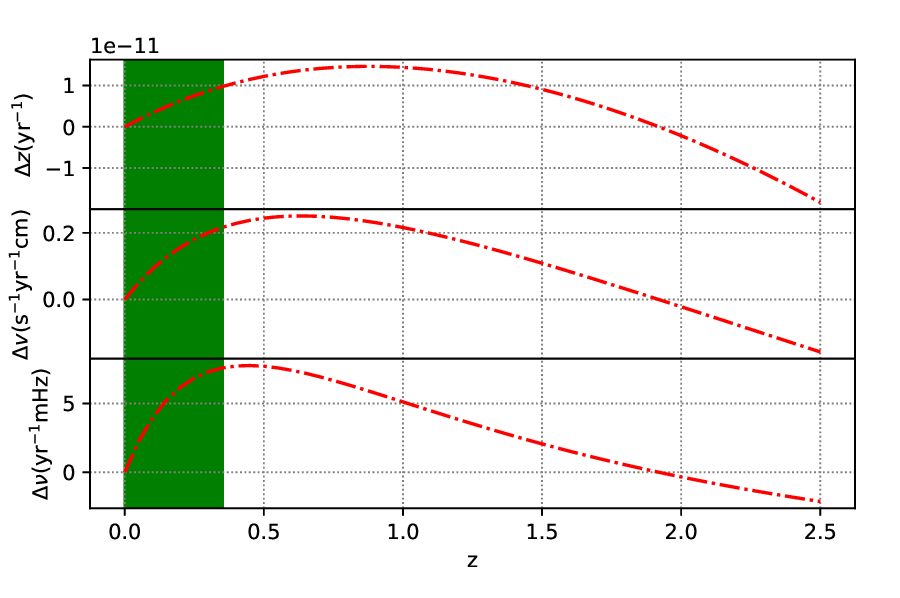}
	\caption{Redshift drift(upper), velocity drift(middle) frequency drift (lower) versus redshift per year. The left green blocks in both panels cover the redshift range of FAST.}
	\label{fig1}
\end{figure}

Applying the inferred late-$\Lambda$CDM-universe parameters\cite{04-AA20-Planck}, $\Omega_{\mathrm{m}0}=0.315,H_0=67.4$ km/s/Mpc, ignoring the slight $\Omega_{\mathrm{r}0}$ and $\Omega_{\mathrm{k}0}$, we can obtain $\Omega_{\mathrm{\Lambda}0}=0.685$ and draw a graph of the mentiond three kinds of drifts versus redshift in Figure \ref{fig1}. Our two observed DLAs both have redshift of 0.2 approximately, the predicted redshift drift $0.6\times10^{-10}\ \mathrm{decade^{-1}}$, velocity drift 1.5 cm/s/decade and frequcency drift 0.06 Hz/decade, declaring a key challenge to resolve at present.

\subsection{Damped Lyman-$\alpha$ absorber} 
Damped Lyman-$\alpha$ Absorber(DLA) is a kind of highly dense HI gas clouds, and most hydrogen in DLA is neutral and empirically have a minimal HI column density threshold $N_{\mathrm{HI}}\approx2\times10^{20}\ \mathrm{cm^{-2}}$\cite{33-MNRAS14-Bird} due to the self-shielding from the ionizing effect of the diffused background radiation. In the targeted radio band, HI 21cm lines result from HI hyperfine structure absorption. Considering the tiny energy difference($0.068k_\mathrm{B}\ \mathrm{J}$) between the two energy levels and much higher spin temperature $T_\mathrm{S}$($100\sim1000\ \mathrm{K}$), the population of HI atoms is relatively stable. In proper DLA 21cm absorption systems with lower temperature and less peculiar motion, HI 21cm absorption would have narrow line width.

Based on the general discussion in Loeb\cite{08-ApJ98-Loeb}, first, the peculiar acceleration of the sun orbiting the galaxy center can amount to cosmic acceleration but be easily removed for its known magnitude and orientation. Second, the effect that typical acceleration inside a virialized object exerts on redshift drift can be small enough.

DLAs can be classified into two types, intervening and associated. The former type usually lies close to a galaxy in the sightline to a radio source, while the latter dwells nearby the source. The associated profiles are broader than intervening ones and have more low optical depth components with high dispersion\cite{34-MNRAS16-Curran}, showing more severe effects from the active areas(such as star-formation or AGN) in the host galaxy. Therefore it is better to observe intervening DLAs.

It is hard to find proper 21cm absorption systems to measure the cosmic acceleration, while it could be remembered that the best constraint at present is given through radio DLA observations using GBT by Darling\cite{26-ApJ12-Darling}. With expectation of the next generation equipments, such as ELT for lyman-$\alpha$ forest method in high redshift($2\sim5$) and full SKA, it is still worth trying with a better radio instrument---FAST to explore latest universe.

\subsection{Observational systematics}\label{sec2.3} 
Relevant systematics can easily affect precise observations. Firstly, several kinds of uncertain magnitudes from peculiar motion are recorded in Bolejko et al.\cite{23-arXiv19-Bolejko}, and we will detail some of them that significantly impact the redshift measurement. The earth's orbital revolution in a half period leads to dipole and quadrupole components, corresponding to redshift changes of $\Delta z_\mathrm{eod}\sim10^{-4}$ and $\Delta z_\mathrm{eoq}\sim10^{-9}$, respectively. The earth's self-rotation in a half period also exerts a  $\Delta z_\mathrm{esr}\sim10^{-6}$. The sun's circling within the Milky Way results in a $\Delta z_\mathrm{so}\sim10^{-10}$, however, we re-examine that, given the cursory LSR speed of 230 km/s, it needs more than 10-million-year revolving to accumulate this redshift difference. Additionally, they mentioned that the tidal and continent motion would both give a $\Delta z_\mathrm{add}\sim10^{-10}$, but we do not find more supporting materials backing it.

Peculiar acceleration would contaminate the measurements as well. The notable one of the sun to the Milky Way, $a_\mathrm{s}$, is inferred to be around 0.8 cm/s/yr in many recent works\cite{24-MNRAS19-Cooke,64-AA18-Titov,65-AA12-Xu}. Admitting that the accurate dynamical modeling of DLAs is still extraordinarily difficult now, Rooke\cite{24-MNRAS19-Cooke} used EAGLE cosmological hydrodynamic simulations to roughly constrain the peculiar acceleration of the cold gas by estimating the gravitational field that may hold the gas. The final result presented the acceleration $a_\mathrm{g}$ of 0.2$\sim$0.4 cm/s/yr in z=0, 1, 2, 3.

Remembering that, in Figure \ref{fig1}, redshift drift and spectroscopic radial velocity drift at z=0.2 are about $0.7\times10^{-11}\mathrm{/yr}$ and $0.15\mathrm{cm/s/yr}$, respectively. The mechanisms of $\Delta z_\mathrm{eod}$, $\Delta z_\mathrm{eoq}$, $\Delta z_\mathrm{esr}$ and $a_\mathrm{s}$ are comparatively clear in theory, however whether the present astrometry could perform such an delicate correction still remains uncertain.

Apart from these complex Doppler effects, one also needs an accurate timing standard to establish the long-term frequency stability of the equipment over one decade. Time should be controlled better than cosmic acceleration with a few nanoseconds, ensuring the accuracy of channel labeling. With the maximum time resolution about 50 $\mathrm{\mu s}$\cite{66-In20-Qian} and the nanosecond-level time control, given the resolution of 100 Hz(33 km/s) near 1140 MHz, FAST would also reach the same precision level in channel labelling($\sim0.3$ cm/s) as ALMA\cite{24-MNRAS19-Cooke}. The level is so similar to the anticipated velocity drift($0.15\mathrm{cm/s/yr}$), and no doubt need subsequent improvement.

Furthermore, due to the lack of detail physical considerations, the given simplified $a_\mathrm{g}$ may only express an ideal case for such measurement. Thus it is vital to observe DLAs, to depict the physical properties(distribution, motion, and inner states) of DLA, and to extend the DLA dataset.

Only close corporation among the astronomical community can reveal these sophisticated mechanisms.

With the previous suggestions, it is explicit for us to select suitable DLAs: (1) be in the observational range of FAST, with declination in $(-14^\circ12',65^\circ48')$, frequency in (1.05,1.42) GHz, redshift in (0,0.3524), representing the lastest accelerating expansion period; (2) be less affected by peculiar motion and far away from the local group, with redshift$>$0.05, while the intervening DLAs and well-observed ones are preferred; (3) have prominent absorption peak and strong continuum, with absorption depth deeper than 0.1 Jy, or optical depth larger than 0.04. We list satisfied DLAs in Table \ref{tab1}, where the few amounts would hamper our observation and analysis.

\begin{table*}[htb]\scriptsize
	\centering
	\begin{tabular}{ccccccccc}
		\hline
		\hline
		backgroud source & RA(J2000) & DEC(J2000) & $z_{\mathrm{DLA}}^c$ & type & $S_{1.4\mathrm{GHz}}^d$ & ref. & opt. depth & FWHM\\
		& (hh mm ss) & (dd mm ss) & & & (Jy) & & & ($\mathrm{km\ s^{-1}}$)\\
		\hline
		\textbf{PKS 0952+179}$^a$ & 09 54 56.824 & +17 43 31.22 & 0.237806 & I & 1.007 & \cite{36-AA01-Kanekar} & 0.013 & 20\\
		\textbf{PKS 1413+135} & 14 15 58.818 & +13 20 23.71 & 0.246079 & A & 1.142 & \cite{37-ApJ92-Carilli} & 0.34 & 18 \\
		\textbf{B2 0738+313 A} & 07 41 10.703 & +31 12 00.23 & 0.091235 & I & 2.051 & \cite{38-ApJ00-Lane} & 0.08 & 13 \\
		\textbf{B2 0738+313 B} & & & 0.220999 & I & & \cite{39-ApJ98-Lane} & 0.042 & 8 \\
		B3 0839+458 & 08 43 07.095 & +45 37 42.90 & 0.192225 & A & 0.259 & \cite{40-AA15-Gereb} & 0.26 & 80 \\
		J1120+2736 & 11 20 30.079 & +27 36 10.83 & 0.111720 & A & 0.152 & \cite{40-AA15-Gereb} & 0.15 & 60 \\
		\emph{J0849+5108}$^b$ & 08 49 57.977 & +51 08 29.02 & 0.311991 & I & 0.344 & \cite{41-AA13-Gupta} & 0.06 & 4 \\
		\emph{J1124+1919} & 11 24 43.693 & +19 19 28.11 & 0.165161 & A & 0.877 & \cite{42-MNRAS06-Gupta} & 0.09 & 13 \\
		\hline
		\hline
	\end{tabular}
	\caption{Selected DLAs in FAST coverage. $^a$ The bold DLAs in the first four lines are observed by Darling \cite{26-ApJ12-Darling}. $^b$ The italic DLAs in the last two lines have obvious multi-peak structure in the reference spectra. $^c$ Redshifts of DLAs in the third column and the type of DLAs in fourth column are given in Curran et al.\cite{34-MNRAS16-Curran}, what we abbreviate are I(intervening) and A(associated). $^d$ The peak flux at 1.4GHz in fifth column refers to the VLA FIRST survey\cite{35-ApJ95-Becker}.The sixth column contains the literature of these original spectra, providing the information in the last two columns.}
	\label{tab1}
\end{table*}

\subsection{Evaluators of the spectra} 
\subsubsection{Signal-to-noise ratio} 
Signal-to-Noise Ratio(SNR) is defined as the signal maximum divided by the signal error, in practice the latter is the root-mean-square value based on the fixed spectral baseline. With System Equivalent Flux Density(SEFD) $SEFD=2k_\mathrm{B}T_\mathrm{sys}A_\mathrm{eff}^{-1}$, the Boltzmann constant $k_\mathrm{B}$ equals\\ $1.386049\times10^{-23}\ \text{J/K}$, FAST L-band receiver has a sensitivity $A_\mathrm{eff}/T_\mathrm{sys}\approx2000\ \text{K/m}^2$ when its system temperature is 20K, hence $SEFD\approx1.7326\times10^{-26}\ \text{J/m}^2=1.7326\ \text{Jy}$. Let us define flux uncertainty $\Delta F=SEFD/\sqrt{\eta_\mathrm{pol}\Delta\nu\Delta t}$, where $\eta$ is polarization, $\Delta \nu$ is frequency resolution, $\Delta t$ is the total integration time of a single DLA, we can deduct the SNR formula theoretically\cite{28-PRL14-YU}:
\begin{equation} 
	SNR=\frac{F}{\Delta F}=\frac{F\sqrt{\eta_\mathrm{pol}\Delta\nu\Delta t}}{SEFD}.
\end{equation}

If we make stringent assumptions that (1) original spectrum has to meet the SNR of 10; (2) frequency resolution needs to reach the level of 0.1 Hz over one decade; (3) the depth of the signal peak is 0.02 Jy; consequently, a 2-polarizations observation for a single DLA requires integration time of $3.839\times10^{7} \text{s}\approx1060\ \text{h}$. But we could add some actual considerations from data processing in the following section that (1) raw spectra often have SNR$<$5, after appropriate processing it can easily exceed 10; (2) using raw spectra to measure cosmic acceleration is out of reality, which necessitates long observations, huge data process, and precise error analysis, so it can be simplified by Gaussian profiles fitting of those processed spectra so we can relax frequency resolution to 10 Hz; (3) narrower and deeper DLA's profiles would be found within FAST coverage, let the absorption depth be 0.05 Jy. As a realistic and achievable medium-term goal, we have a more friendly time 3600 s = 1 h.

\subsubsection{Uncertainty of velocity measurement}\label{sec2.4.2}
Fitting Gaussian profiles to extract spectral features, where we use the mean to denote cosmic acceleration peak location and the standard deviation to depict the narrowness of the profile, proposes a new question: what uncertainty the fitting can give, or how many significant digits the results would present. In this paper, we introduce three indicators as temporary solutions and employ them simultaneously.

We regard the result of Fouque et al.\cite{44-AAS90-Fouque} as a more reasonable way, in which they theoretically deducted the velocity uncertainty of a Gaussian profile considering as more factors, satisfying our case where using Gaussian profile to fit the line is necessary. We rewrite their final formula as follow:
\begin{equation} 
	\sigma_{\mathrm{V}1}=0.797885\frac{\sqrt{RA}}{S\sqrt{h}}\ (\mathrm{km\ s^{-1}}),
\end{equation}
where $R$ is velocity resolution(km/s), $A$ is the area enveloped by the profile and coordinate axis(Jy km/s), $S$ is the SNR, $h$ is the peak signal(Jy).

Meanwhile, as a more used and semi-empirical result, we use what Koribalski et al.\cite{45-AJ04-Kori} proposed and Zhang et al.\cite{46-MNRAS21-ZHANG} modified for FAST as an alternative:
\begin{equation} 
	\sigma_{\mathrm{V}2}=3\frac{\sqrt{PR}}{S}\ (\mathrm{km\ s^{-1}}),
\end{equation}
where $R$ is velocity resolution(km/s), $w_{20}$ indicates the profile width at 20\% peak value, $P=0.5(w_{20}-w_{50})$ is defined to express the slope of the profile, $S$ is the SNR.

We notice that $\sigma_{\mathrm{V}1}$ and $\sigma_{\mathrm{V}2}$ are theoretical estimators, and need to find one from data directly. However with little data, it is unreliable to establish a complex connection between data features and velocity uncertainty. Given that Gauss profile is a probability density function, we can define a velocity width $\sigma_{\mathrm{V}3}$ as a stopgap that it takes up the 1\% of whole probability and is symmetric along the mean-velocity axis. It is clear that the new indicator only relates to ideal function parameters and is independent of detailed data features such as SNR and resolution.

\section{Observations and data} 
\subsection{Observational settings}\label{sec3.1}
In the FAST commissioning progress\cite{47-SCPMA19-JIANG}, we use its 19-beams L-band receiver to observe two DLAs(PKS0952+179, PKS1413+135) on Nov. 17th, 19th, 2019 respectively and collect the baseband data of the central beam covering 1.00-1.50 GHz. 15-minute ON-OFF mode is applied for each source with 10 K high noise injected every other second approximately, two polarizations recorded, and 8-bit sampling.

It is our first time implementing data processing, so we prepare a set of Fast Fourier Transformed data of 16sec baseband from PKS1413(TEST data). Some data at the beginning of observation are easily affected by mechanically controlled motion, while some data of PKS0952 corrupt. Eliminating those damaged data, we confirm that in the left 10-minutes FFT data of PKS1413 with 10 Hz resolution, the first half is in source-ON mode while the last is in the OFF mode. We use the TEST data at first, then employ the tested methods to process the REAL data.

\subsection{Data process}\label{sec3.2}
\begin{figure}[htb]
	\centering
	\includegraphics[scale=0.6]{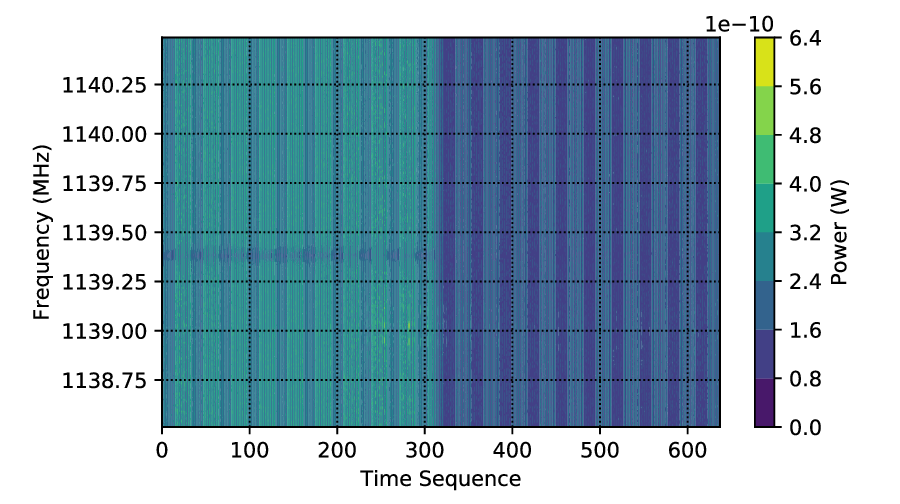}
	\caption{The waterfall of REAL data(PKS1413) with about 10 kHz resolution, the abscissa is a time sequence but may be discontinuous. With narrow vertical stripes suggesting diode states, the left pale portion is source-ON, and the right dark is source-OFF. We can find a faint absorption line near 1139 MHz in the left part. In diode stripes there exist narrower dark vertical stripes explained in Figure \ref{fig4}.}
	\label{fig2}
\end{figure}

We cut off the 10 MHz-width band containing the expected absorption from the 500MHz baseband data and draw its waterfall to roughly judge whether we detect the absorption peak, 1142.5-1152.5MHz for PKS0952 and 1135.0-1145.0MHz for PKS1413. In the waterfalls of PKS 1413 we can see a clear horizontal absorption line near 1139.4MHz (Figure \ref{fig2}), while no evident absorption is observed at the waterfall of PKS0952.

\begin{figure}[htb]
	\centering
	\includegraphics[scale=0.6]{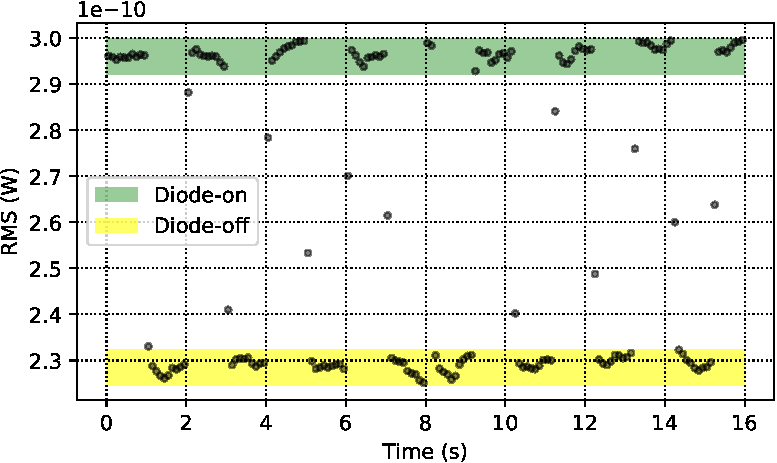}
	\caption{The RMS noise behavior of TEST data(only including source-ON data). The diode effect is distinguishable in the time direction. The higher data points covered by the green block represent diode-on, while the lower in the yellow block is diode-off.}
	\label{fig3}
\end{figure}

\begin{figure}[htb]
	\centering
	\includegraphics[scale=0.6]{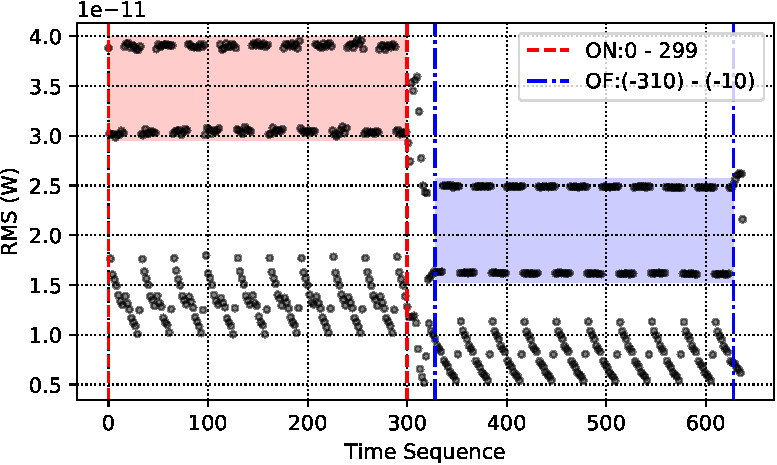}
	\caption{The RMS noise behavior of REAL data. At the upper part of the graph, we can see two arrays in the color blocks with a similar pattern to Figure \ref{fig3}. But at the lower part, noise behavior is different, descending periodically, covering all the observational time, and indicating some systematics in the instruments.}
	\label{fig4}
\end{figure}

Then we divide the states of source-ON/OFF and diode-on/off by analysing their RMS noise level free from the absorption and RFI. From TEST data including source-ON only(Figure \ref{fig3}), we find distinct regularity of the diode modes in the shape of high and low impulses. As for REAL data(Figure \ref{fig4}), we utilize the same regularity to classify the upper part data as on/off, while the left higher part denotes source-ON mode. Curiously, at the lower part of Figure \ref{fig4}, the rms values align in a totally different arrangement to their upper counterparts, thus are abandoned from date process out of caution and processed in the next subsection.

The next step is calibrating temperature spectra from power ones. At first, convert them into source-ON/OFF system-temperature spectra individually:
\begin{eqnarray}
	&T_1=&T_\mathrm{sys}^\mathrm{on}=\frac{P^\mathrm{on}}{P^\mathrm{on}-P^\mathrm{off}}T_\mathrm{cal}-T_\mathrm{cal},\\
	&T_2=&T_\mathrm{sys}^\mathrm{off}=\frac{P^\mathrm{off}}{P^\mathrm{on}-P^\mathrm{off}}T_\mathrm{cal},
\end{eqnarray}
where $T_\mathrm{cal}$ is the diode temperature($\approx$10 K). To reduce the glitch-interference in baseline fitting require data smoothing. Avoiding flattening the peaks, we refer to Zhang et al. \cite{46-MNRAS21-ZHANG} and only make Gaussian smooth of the denominator every 0.5 MHz considering Nyquist sampling and 1 MHz baseline ripple of the FAST receiver system.

With a reasonable assumption that similar RMS noise levels exist in the divided two sets of data(source-ON and -OFF), we obtain an RMS($\sigma$) weighted spectrum to minimize the merged RMS noise level:
\begin{eqnarray}
	& &w_\mathrm{i}=\sigma_\mathrm{j}^2/(\sigma_1^2+\sigma_2^2),\ i,j=1,2,\ i\neq j,\\
	& &T_\mathrm{sys}=w_1T_1+w_2T_2.
\end{eqnarray}

\begin{figure}[tbp]
	\centering
	\includegraphics[scale=0.6]{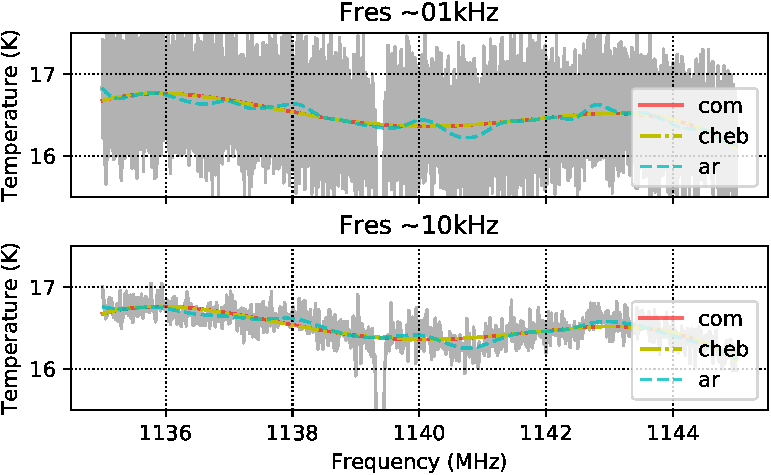}
	\caption{Baseline fittings of REAL data with 10 MHz-band in the low-res group. The red curve is cubic polynomial plus sine fitting, the yellow dot line is Chebyshev fitting, and the cyan dashed line is arPLS fitting. In the 10 kHz case, the arPLS is comparable to the others, but it can extract finer variances in the resolution in the high resolution up to 1 kHz.}
	\label{fig5}
\end{figure}

\begin{figure}[tbp]
	\centering
	\includegraphics[scale=0.6]{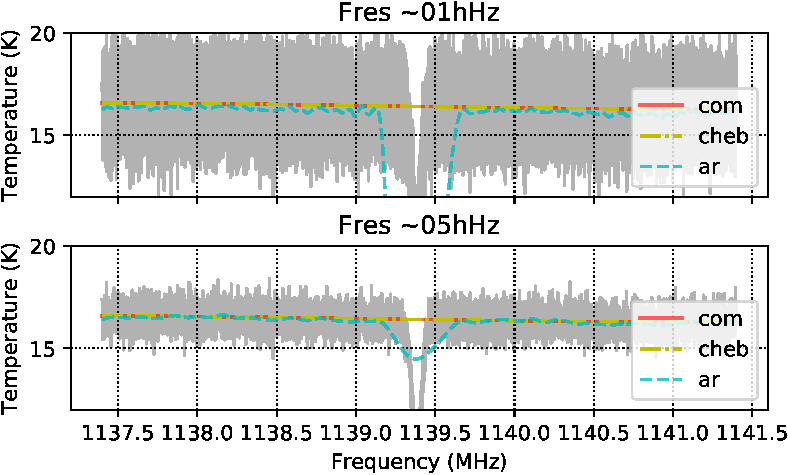}
	\caption{Baseline fittings of REAL data with 4 MHz-band in the high-res group. The arPLS is unfit to express the baseline anymore.}
	\label{fig6}
\end{figure}

For TEST data only has ON spectrum, we will concern NO-OFF fitting of REAL data. We compare 3 baseline fittings: (1) cubic polynomial plus sine(com), (2) Chebyshev polynomial(cheb), (3) asymmetric reweighted Penalty Least Square(arPLS, Beak et al.\cite{48-Analyst15-Beak}) in the HiFAST pipeline(Jing\footnote{Yingjie Jing: \url{jyj@nao.cas.cn}} et al. preparation). In a low resolution(10kHz), three means have comparable results in RMS noise, while at higher resolution(5 kHz, 2 kHz, 1 kHz), the arPLS surpasses the others(Figure \ref{fig5}). So we use arPLS fitting in a low-res group(from 10 kHz to 1 kHz). However, in a finer group(5 hundredHz, 2 hHz, 1 hHz), the arPLS is abnormal around the peak due to the noticeable sensitivity and continuous differentiability(Figure \ref{fig6}). Therefore we use cheb fitting in the high-res group(higher than 1 kHz). After baseline correction, we divided by the antenna gain of G(1150 MHz)=16.48 K/Jy\cite{49-RAA20-JIANG} around the targeted frequency and cut 1 MHz-width flux density-frequency data for the last Gaussian fitting.

\begin{figure}[tbp]
	\centering
	\includegraphics[scale=0.6]{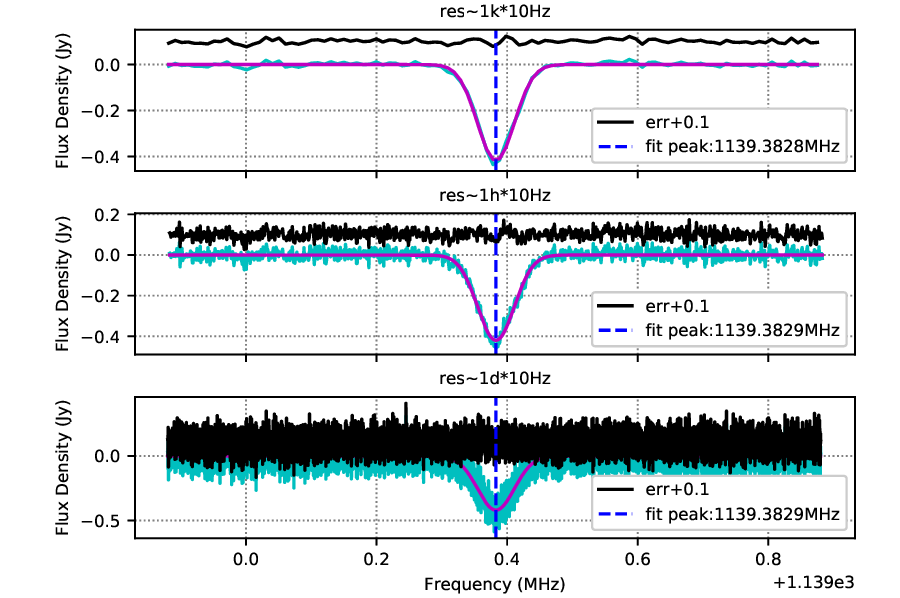}
	\caption{Frequency Gaussian fittings of REAL data with 1 MHz-band. There are $\approx$10, 1, 0.1 kHz resolutions from top to bottom. The cyan line is realistic data, the magenta curve is the Gaussian profile we fit, the black line is the offseted error of Gaussian fitting. The vertical blue dashed line is the fitted peak location.}
	\label{fig7}
\end{figure}

\begin{figure}[tbp]
	\centering
	\includegraphics[scale=0.6]{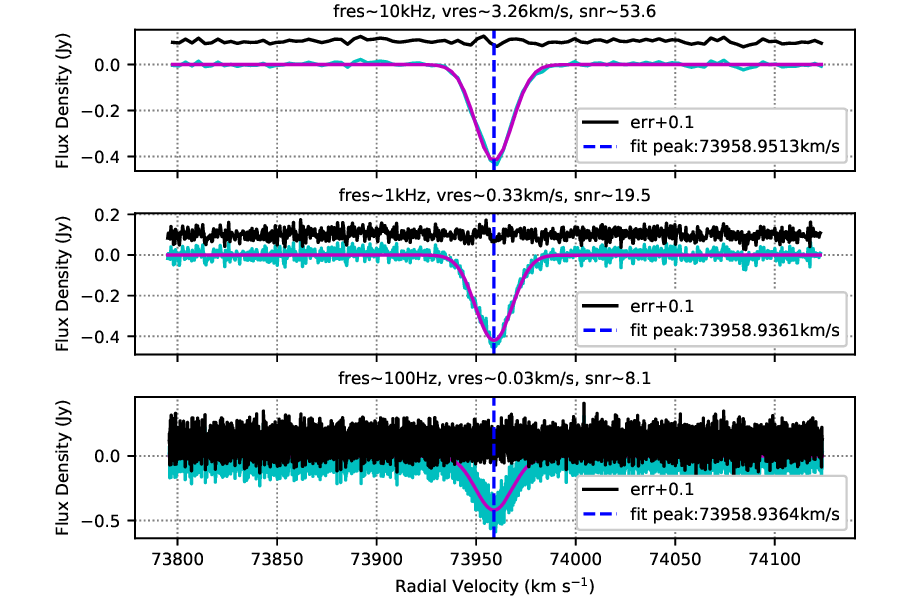}
	\caption{Velocity Gaussian fittings of REAL data with 1 MHz-band. There are $\approx$3.26, 0.33, 0.03 $\mathrm{km\ s^{-1}}$ resolutions from top to bottom.}
	\label{fig8}
\end{figure}

We make a frequency Gaussian fitting to the truncated data(Figure \ref{fig7}) with a custom Gaussian function and the optimize.curve\_fit function in the python3 package scipy. Transforming the frequency to barycentric velocity via PyAstronomy module\cite{50-ascl2019-Czesla} under the online instruction(the function: radial\_velocity\_correction\footnote{\url{https://docs.astropy.org/en/stable/api/astropy.coordinates.SkyCoord.html}}) of the astropy module\cite{51-AA2013-astropy}, ultimately, we take a velocity Gaussian fitting(Figure \ref{fig8}) and analyze the results by the former two indicators in section \ref{sec2.4.2}.

\subsection{Results}\label{sec3.3}
We cut off some bands of different widths to fit the baseline with several frequency resolutions but always fit the Gaussian profile in 1.0 MHz band, and list some essential fitted parameters in Table \ref{tab2}. The consistency of fitted central frequency and radial velocity are manifested remarkably in all the cases. To guarantee a persuasive result, we ignore the data with SNR$<$15 in the following discussion.

\begin{table*}[tbp]\scriptsize
	\centering
	\begin{tabular}{ccccccccc}
		\hline
		\hline
		Band width$^a$ & SNR & $\mathrm{RES_\mathrm{\nu}}$ & $\mathrm{RES_V}$ & Fitted $\mu^b$ & Fitted $\sigma^b$ & $\sigma_{\mathrm{V}1}$ & $\sigma_{\mathrm{V}2}$ & $\sigma_{\mathrm{V}3}$ \\
		(MHz) & & (kHz) & ($\mathrm{km\ s^{-1}}$) & ($\mathrm{km\ s^{-1}}$) & ($\mathrm{km\ s^{-1}}$) & ($\mathrm{km\ s^{-1}}$) & ($\mathrm{km\ s^{-1}}$) & ($\mathrm{km\ s^{-1}}$) \\
		\hline
		1135.0$\sim$1145.0 & 20.40 & $\sim1$ & 0.326  & 73958.9342 & 9.2912 & 0.1078 & 0.2010 & 0.1165\\
		& 52.80 & $\sim10$ & 3.259  & 73958.9426 & 9.4310 & 0.1326 & 0.2474 & 0.1182 \\
		1137.4$\sim$1141.4 & 8.02 & $\sim0.1$ & 0.033  & 73958.9449 & 9.4320 & 0.0873 & 0.1629 & 0.1182\\
		& 20.38 & $\sim1$ & 0.326  & 73958.9356 & 9.3068 & 0.1079 & 0.2013 & 0.1166\\
		& 55.43 & $\sim10$ & 3.259  & 73958.9367 & 9.4725 & 0.1266 & 0.2361 & 0.1187\\
		1138.4$\sim$1140.4 & 8.17 & $\sim0.1$ & 0.033  & 73958.9361 & 9.4030 & 0.0856 & 0.1697 & 0.1179\\
		& 19.91 & $\sim1$ & 0.326  & 73958.9377 & 9.3394 & 0.1106 & 0.2064 & 0.1171\\
		& 52.78 & $\sim10$ & 3.259  & 73958.9548 & 9.4483 & 0.1328 & 0.2477 & 0.1184\\
		\hline
		\hline
	\end{tabular}
	\caption{Key fitted parameters from the DLA in PKS1413+135 spectra. \textit{Notes:} $^a$ The diverse cut-bands are only used in baseline correction. $^b$ The columns from the fifth to seventh are the three parameters of a Gaussian profile $f(x)=\frac{\alpha}{\sqrt{2\pi}\sigma}\exp{\frac{(x-\mu)^2}{2\sigma^2}}$. The fifth is the fit barycentric radial velocity, and the sixth is the fit standard deviation. All fitted $\alpha$ are in (-0.424, -0.421) Jy, consequently we do not present them.}
	\label{tab2}
\end{table*}

Various resolutions smooth distinct data points, different wide-bands retain unequal data amount, both may slightly change the data distribution and naturally make the results fluctuate. Every 0.001 km/s is neglectable in common analysis but precious in cosmic acceleration scenario.

Un-independent spectra and their consistency make the producing of final result difficult. Here we take the smallest $\sigma_{\mathrm{V}1}$ one(with SNR higher than 15) and will discuss later: the corrected barycentric radial velocity $V_\mathrm{M}=73958.9356\pm0.1079$ km/s, corrected redshift $z_\mathrm{M}$=0.246700 45(36), corrected frequency $f_\mathrm{M}=1139332.02\pm0.33$ kHz.

\subsection{Discussions}\label{sec3.4}
As for the absent absorption features of PKS0952, we find a reason: its absorption depth is less than 0.02 Jy obtained by 5 h source-ON integration\cite{36-AA01-Kanekar}. We have only about 5 min ON time, and 2/3 of the data are accidentally damaged.

Referring to the literature spectrum of PKS1413\cite{37-ApJ92-Carilli}, we reach a stricter redshift constraint(where it gave $z=0.24671\pm0.00001$), a more Gaussian profile, and a less notable extended wing. According to Darling \cite{26-ApJ12-Darling}, the DLA in the PKS1413 spectrum should be a steady one free of noticeable frequency shift with redshift $z_\mathrm{HI}=0.24670374(30)$ based on a decade observation. However, in Curran et al.\cite{34-MNRAS16-Curran}, it has the redshift of $z_\mathrm{HI}=0.246079$. A 0.0006 difference exists in the two redshift values, and our result supports Darling's redshift but still have a redshift difference. Although Darling claimed the frequency stability of DLA in PKS1413 by the long-span observations, we notice that Curran classified it as an associated one, which is more likely to be affected by the host galaxy. Consequently, the difference in redshifts is reasonable, justifying the significance of persistent concern about DLAs, and requires more observations to explain. Restricted by the shortage of observation time, at a similar resolution(10 kHz), we only achieve an RMS noise of 0.008 Jy, worse than the literature. But all of these finished in a twelve-minute observation already proved the great power of FAST.

Comparing the error bar of redshift we obtain, $\Delta z_\mathrm{F}\sim10^{-7}$, with those introduced redshift uncertainties in section \ref{sec2.3}, we find the precision performs much better than the dipole component of the earth's orbital revolution ($\Delta z_\mathrm{eod}$ $\sim10^{-4}$), a little higher than the earth's self-rotation ($\Delta z_\mathrm{esr}$ $\sim10^{-6}$). The performance is quite natural because the time interval in observation is much less than a half day, let alone a half year, and can prove the precision of FAST and our pipeline preliminarily. Nevertheless, $\Delta z_\mathrm{F}$ is still far away to the quadrupole component of the earth's orbital revolution($\Delta z_\mathrm{eoq}\sim10^{-9}$), and to those more precise ones($\lesssim10^{-10}$), showing the great potential as well as challenge of the astrometry.

Comparing the former two indicators of velocity uncertainty, we find that they all decrease when resolution improves, matching our presentative assumptions. $\sigma_{\mathrm{V}2}$ is almost one time larger than $\sigma_{\mathrm{V}1}$, which we consider as more reasonable gives a stricter constraint, for its derivation is more theoretical and less affected by geometirc and numerical process, while it still contains same spectral information such as SNR and resolution. The two evaluators originate from certain features in spectra and ignore the undergoing physics and systematics discussed in Liske et al. \cite{25-MNRAS08-Liske}. $\sigma_{\mathrm{V}3}$ keeps constant under consistent parameters, and is more closer to $\sigma_{\mathrm{V}1}$ in every case supporting our decision to adopt $\sigma_{\mathrm{V}1}$, while we must remember that $\sigma_{\mathrm{V}3}$ is a temporary tool and even captures much less data details. Due to these drawbacks, we use them especially the well-performed $\sigma_{\mathrm{V}1}$ out of expediency and still need to find a better uncertainty standard with more clear physics and data considerations.

Aiming to trace cosmic acceleration, the most serious question we encounter is how to improve spectral resolution and decrease the velocity uncertainty at the same time. High-resolved data demand more observation time, while the current uncertainty does not halve when the resolution improves by two orders of magnitude. It is necessary to prolong the integration time per DLA because the 10 min is too short and can not help us reach the limitation of velocity uncertainty that FAST baseband data can offer.

The biggest technical problem is the process of massive data. Though a 10 MHz band would satisfy our need, the FAST baseband provides 500 MHz data before Fourier Transform, raising a rigorous demand on the computational capacity in the FAST site and the potential optimization of our FFT program.

Without adequate integration time, the fluctuations in lines rise with the increase of resolution. In this case, com and cheb are not good, while arPLS is invalid for its abnormal behavior near the peak. We should update the model parameters in arPLS at the different scenarios or find a finer baseline fitting. The new method must take apart global subtle baseline variations from local trivial and sharp interference, where their boundary gradually blurs.

We perform barycentric radial velocity correction via PyAstronomy and astropy with the same input. But the outputs they produce have a difference of about 6 m/s. We adopt the PyAstronomy outcome for it can present detailed information which can be verified conveniently. Though the uncertainty we obtain is one order of magnitude larger than 6 m/s, more accurate astrometric techniques are still in need.

Moreover, the values we process in a python3 program are operated in the form of floating-point numbers. With its accuracy to fifteen decimal places, this data type could cause the stochastic fluctuating in intermediate results, which would lead to a wrong outcome, especially in a high-precision case. The round function or decimal module both are unfit to process massive data in an array or matrix.

For the first attempt, we have been learning from some flaws. The most irreparable is that we document the sequence numbers of the FFT data but neglect exact time information. After removing corrupted data, the timeline becomes discontinuous, preventing us from improving the precision of radial velocity correction. Besides, since our critical concern is peak location(velocity), we should use more diode-off and less source-OFF time during the observation. 

Approaching radio cosmic acceleration raises some crucial necessities. (1) An accurate barycentric radial velocity correction with the accuracy of cm/s/decade. Now the results given by PyAstronomy and astropy still have a difference about 6 m/s. (2) A precise timing standard to establish and verify the long-term frequency stability of the telescope system and DLAs at the same time. (3) A statistical sample set of DLAs diminishing the bias. (4) A recognized and analytical velocity uncertainty to show us the more realistic numerical gap between theory and observation. With intimate and comprehensive collaboration across the community, cosmic acceleration and SL cosmology will be no longer far away.

For the final scientific goal of cosmic acceleration signal arising from consecutive observatios, we are persistenlty applying for more observing time in FAST and updating our data processing flow with adpative methods. With the improvement in the maximal spectral resolution of FAST(SERENDIP VI system\cite{57-ApJ20-Zhang} designed for SETI in FAST can produce spectral resolution of 3.725Hz, and the last SETI observation in FAST(Tao\footnote{Zhenzhao Tao: \url{tzz@mail.bnu.edu.cn}} et al. preparation) realize frequence channel of 7.5Hz in practice), more proper DLAs verified(we also notice that the pilot extragalactic HI survey in FAST just publish its first data release\cite{58-arXiv22-Kang}, and we hope such projects can detect more DLAs in FAST coverage) and the introduction of adaptive technologies, we aim to examine the technical feasibility and expect to reach our goal in a decade, boosting our understanding in late-time universe expansion and mysterious dark energy.

FAST observations will cover most northern hemisphere and low redshift(0 $\sim$ 0.35) regions, and are perfect complement with the planned ground-based lyman-$\alpha$ forest projects(such as ELT) which more focus on southern hemisphere or higher redshift(1.5 $\sim$ 5) region. Additionally the ongoing First Large Absorption Survey in HI(FLASH) with ASKAP\cite{56-PASA22-Allison} would provide more HI absorption systems in redshift range 0.4 $\sim$ 1 for future similiar radio observations in SKA. All these new generation telescopes are widely expected to obtain more precise results.

\section{Conclusion} \label{sec4}
By observing two DLAs in the FAST commissioning progress, we manage to attain an HI 21cm absorption line feature and prove the consistency of Gaussian fitting results at different baseline lengths and resolutions, validating the correctness of our processing. Constrained by the observation time, the spectra we generated have a deficiency in RMS noise level than the literature spectrum\cite{37-ApJ92-Carilli}, but have higher precision and better Gaussian shape, providing convincing feasibility of Gaussian fitting. Under the demand for precise velocity measurement, we establish an analytical framework two phenomenological evaluators from spectral features to represent velocity uncertainty.

We take these final parameters of DLA in the sightline of PKS1413+135: the corrected barycentric radial velocity $V_\mathrm{M}=73958.9356\pm0.1079$ km/s, corrected redshift $z_\mathrm{M}=0.24670045\pm0.00000036$, corrected frequency $f_\mathrm{M}=1139332.02\pm0.33\ \mathrm{kHz}$. The redshift we derive is more precise than the original literature\cite{37-ApJ92-Carilli}, but a bit different with Darling \cite{26-ApJ12-Darling}. It is a good constraint based on the detailed conditions we have, but still far away from constraining a persuasive cosmic acceleration signal and needs more observing. Therefore it is vital and worthy to advance the observations, to make full use of FAST, and to answer whether it could be used for the final goal---detecting cosmic acceleration. Our FAST attempt could have the technical ability to conduct redshift drift experiments practically, and we will continue to research the possibility in radio DLA approach.

\section{Acknowledgements} \label{sec5}
We thank the anonymous referee for the kind comments that helped us greatly to improve this paper. We thank Jie-Feng Chen, Wenkai Hu, Di Li, Ue-Li Pen, Yougang Wang, Zijian Wang, Zhong-Zu Wu, Yidong Xu, You-Ling Yue, and Zheng Zheng for useful discussions. We particullarly thank Yingjie Jing for the HiFAST pipeline processing HI data from the FAST, Ningyu Tang for the the assumption of same RMS in temperature calibration, and J. Darling for the generous discussion of the absorption line for PKS1413. This work is supported by the National Natural Science Foundation of China(Grants No. 61802428,11929301) and the National Key R\&D Program of China(2017YFA0402600).

\bibliographystyle{elsarticle-num-names}
\bibliography{lcz-v1.6}

\begin{thebibliography}{64}
\expandafter\ifx\csname natexlab\endcsname\relax\def\natexlab#1{#1}\fi
\providecommand{\url}[1]{\texttt{#1}}
\providecommand{\href}[2]{#2}
\providecommand{\path}[1]{#1}
\providecommand{\DOIprefix}{doi:}
\providecommand{\ArXivprefix}{arXiv:}
\providecommand{\URLprefix}{URL: }
\providecommand{\Pubmedprefix}{pmid:}
\providecommand{\doi}[1]{\href{http://dx.doi.org/#1}{\path{#1}}}
\providecommand{\Pubmed}[1]{\href{pmid:#1}{\path{#1}}}
\providecommand{\bibinfo}[2]{#2}
\ifx\xfnm\relax \def\xfnm[#1]{\unskip,\space#1}\fi
\bibitem[{{Riess} et~al.(1998){Riess}, {Filippenko}, {Challis}, {Clocchiatti},
  {Diercks}, {Garnavich}, {Gilliland}, {Hogan}, {Jha}, {Kirshner},
  {Leibundgut}, {Phillips}, {Reiss}, {Schmidt}, {Schommer}, {Smith},
  {Spyromilio}, {Stubbs}, {Suntzeff}, and {Tonry}}]{01-AJ98-Riess}
\bibinfo{author}{A.~G. {Riess}}, \bibinfo{author}{A.~V. {Filippenko}},
  \bibinfo{author}{P.~{Challis}}, \bibinfo{author}{A.~{Clocchiatti}},
  \bibinfo{author}{A.~{Diercks}}, \bibinfo{author}{P.~M. {Garnavich}},
  \bibinfo{author}{R.~L. {Gilliland}}, \bibinfo{author}{C.~J. {Hogan}},
  \bibinfo{author}{S.~{Jha}}, \bibinfo{author}{R.~P. {Kirshner}},
  \bibinfo{author}{B.~{Leibundgut}}, \bibinfo{author}{M.~M. {Phillips}},
  \bibinfo{author}{D.~{Reiss}}, \bibinfo{author}{B.~P. {Schmidt}},
  \bibinfo{author}{R.~A. {Schommer}}, \bibinfo{author}{R.~C. {Smith}},
  \bibinfo{author}{J.~{Spyromilio}}, \bibinfo{author}{C.~{Stubbs}},
  \bibinfo{author}{N.~B. {Suntzeff}}, \bibinfo{author}{J.~{Tonry}},
\newblock \bibinfo{title}{{Observational Evidence from Supernovae for an
  Accelerating Universe and a Cosmological Constant}},
\newblock \bibinfo{journal}{\aj} \bibinfo{volume}{116} (\bibinfo{year}{1998})
  \bibinfo{pages}{1009--1038}. \DOIprefix\doi{10.1086/300499}.
  \href{http://arxiv.org/abs/astro-ph/9805201}{{\tt arXiv:astro-ph/9805201}}.
\bibitem[{{Perlmutter} et~al.(1999){Perlmutter}, {Aldering}, {Goldhaber},
  {Knop}, {Nugent}, {Castro}, {Deustua}, {Fabbro}, {Goobar}, {Groom}, {Hook},
  {Kim}, {Kim}, {Lee}, {Nunes}, {Pain}, {Pennypacker}, {Quimby}, {Lidman},
  {Ellis}, {Irwin}, {McMahon}, {Ruiz-Lapuente}, {Walton}, {Schaefer}, {Boyle},
  {Filippenko}, {Matheson}, {Fruchter}, {Panagia}, {Newberg}, {Couch}, and
  {Project}}]{02-ApJ99-Perlmutter}
\bibinfo{author}{S.~{Perlmutter}}, \bibinfo{author}{G.~{Aldering}},
  \bibinfo{author}{G.~{Goldhaber}}, \bibinfo{author}{R.~A. {Knop}},
  \bibinfo{author}{P.~{Nugent}}, \bibinfo{author}{P.~G. {Castro}},
  \bibinfo{author}{S.~{Deustua}}, \bibinfo{author}{S.~{Fabbro}},
  \bibinfo{author}{A.~{Goobar}}, \bibinfo{author}{D.~E. {Groom}},
  \bibinfo{author}{I.~M. {Hook}}, \bibinfo{author}{A.~G. {Kim}},
  \bibinfo{author}{M.~Y. {Kim}}, \bibinfo{author}{J.~C. {Lee}},
  \bibinfo{author}{N.~J. {Nunes}}, \bibinfo{author}{R.~{Pain}},
  \bibinfo{author}{C.~R. {Pennypacker}}, \bibinfo{author}{R.~{Quimby}},
  \bibinfo{author}{C.~{Lidman}}, \bibinfo{author}{R.~S. {Ellis}},
  \bibinfo{author}{M.~{Irwin}}, \bibinfo{author}{R.~G. {McMahon}},
  \bibinfo{author}{P.~{Ruiz-Lapuente}}, \bibinfo{author}{N.~{Walton}},
  \bibinfo{author}{B.~{Schaefer}}, \bibinfo{author}{B.~J. {Boyle}},
  \bibinfo{author}{A.~V. {Filippenko}}, \bibinfo{author}{T.~{Matheson}},
  \bibinfo{author}{A.~S. {Fruchter}}, \bibinfo{author}{N.~{Panagia}},
  \bibinfo{author}{H.~J.~M. {Newberg}}, \bibinfo{author}{W.~J. {Couch}},
  \bibinfo{author}{T.~S.~C. {Project}},
\newblock \bibinfo{title}{{Measurements of {\ensuremath{\Omega}} and
  {\ensuremath{\Lambda}} from 42 High-Redshift Supernovae}},
\newblock \bibinfo{journal}{\apj} \bibinfo{volume}{517} (\bibinfo{year}{1999})
  \bibinfo{pages}{565--586}. \DOIprefix\doi{10.1086/307221}.
  \href{http://arxiv.org/abs/astro-ph/9812133}{{\tt arXiv:astro-ph/9812133}}.
\bibitem[{{Scolnic} et~al.(2018){Scolnic}, {Jones}, {Rest}, {Pan}, {Chornock},
  {Foley}, {Huber}, {Kessler}, {Narayan}, {Riess}, {Rodney}, {Berger}, {Brout},
  {Challis}, {Drout}, {Finkbeiner}, {Lunnan}, {Kirshner}, {Sanders},
  {Schlafly}, {Smartt}, {Stubbs}, {Tonry}, {Wood-Vasey}, {Foley}, {Hand},
  {Johnson}, {Burgett}, {Chambers}, {Draper}, {Hodapp}, {Kaiser}, {Kudritzki},
  {Magnier}, {Metcalfe}, {Bresolin}, {Gall}, {Kotak}, {McCrum}, and
  {Smith}}]{03-ApJ18-Scolnic}
\bibinfo{author}{D.~M. {Scolnic}}, \bibinfo{author}{D.~O. {Jones}},
  \bibinfo{author}{A.~{Rest}}, \bibinfo{author}{Y.~C. {Pan}},
  \bibinfo{author}{R.~{Chornock}}, \bibinfo{author}{R.~J. {Foley}},
  \bibinfo{author}{M.~E. {Huber}}, \bibinfo{author}{R.~{Kessler}},
  \bibinfo{author}{G.~{Narayan}}, \bibinfo{author}{A.~G. {Riess}},
  \bibinfo{author}{S.~{Rodney}}, \bibinfo{author}{E.~{Berger}},
  \bibinfo{author}{D.~J. {Brout}}, \bibinfo{author}{P.~J. {Challis}},
  \bibinfo{author}{M.~{Drout}}, \bibinfo{author}{D.~{Finkbeiner}},
  \bibinfo{author}{R.~{Lunnan}}, \bibinfo{author}{R.~P. {Kirshner}},
  \bibinfo{author}{N.~E. {Sanders}}, \bibinfo{author}{E.~{Schlafly}},
  \bibinfo{author}{S.~{Smartt}}, \bibinfo{author}{C.~W. {Stubbs}},
  \bibinfo{author}{J.~{Tonry}}, \bibinfo{author}{W.~M. {Wood-Vasey}},
  \bibinfo{author}{M.~{Foley}}, \bibinfo{author}{J.~{Hand}},
  \bibinfo{author}{E.~{Johnson}}, \bibinfo{author}{W.~S. {Burgett}},
  \bibinfo{author}{K.~C. {Chambers}}, \bibinfo{author}{P.~W. {Draper}},
  \bibinfo{author}{K.~W. {Hodapp}}, \bibinfo{author}{N.~{Kaiser}},
  \bibinfo{author}{R.~P. {Kudritzki}}, \bibinfo{author}{E.~A. {Magnier}},
  \bibinfo{author}{N.~{Metcalfe}}, \bibinfo{author}{F.~{Bresolin}},
  \bibinfo{author}{E.~{Gall}}, \bibinfo{author}{R.~{Kotak}},
  \bibinfo{author}{M.~{McCrum}}, \bibinfo{author}{K.~W. {Smith}},
\newblock \bibinfo{title}{{The Complete Light-curve Sample of Spectroscopically
  Confirmed SNe Ia from Pan-STARRS1 and Cosmological Constraints from the
  Combined Pantheon Sample}},
\newblock \bibinfo{journal}{\apj} \bibinfo{volume}{859} (\bibinfo{year}{2018})
  \bibinfo{pages}{101}. \DOIprefix\doi{10.3847/1538-4357/aab9bb}.
  \href{http://arxiv.org/abs/1710.00845}{{\tt arXiv:1710.00845}}.
\bibitem[{{Planck Collaboration}(2020)}]{04-AA20-Planck}
\bibinfo{author}{{Planck Collaboration}},
\newblock \bibinfo{title}{{Planck 2018 results. VI. Cosmological parameters}},
\newblock \bibinfo{journal}{\aap} \bibinfo{volume}{641} (\bibinfo{year}{2020})
  \bibinfo{pages}{A6}. \DOIprefix\doi{10.1051/0004-6361/201833910}.
  \href{http://arxiv.org/abs/1807.06209}{{\tt arXiv:1807.06209}}.
\bibitem[{{Hong} et~al.(2016){Hong}, {Han}, and {Wen}}]{05-ApJ16-HONG}
\bibinfo{author}{T.~{Hong}}, \bibinfo{author}{J.~L. {Han}},
  \bibinfo{author}{Z.~L. {Wen}},
\newblock \bibinfo{title}{{A Detection of Baryon Acoustic Oscillations from the
  Distribution of Galaxy Clusters}},
\newblock \bibinfo{journal}{\apj} \bibinfo{volume}{826} (\bibinfo{year}{2016})
  \bibinfo{pages}{154}. \DOIprefix\doi{10.3847/0004-637X/826/2/154}.
  \href{http://arxiv.org/abs/1511.00392}{{\tt arXiv:1511.00392}}.
\bibitem[{{Chen} et~al.(2018){Chen}, {Fishbach}, and {Holz}}]{06-Nature18-Chen}
\bibinfo{author}{H.-Y. {Chen}}, \bibinfo{author}{M.~{Fishbach}},
  \bibinfo{author}{D.~E. {Holz}},
\newblock \bibinfo{title}{{A two per cent Hubble constant measurement from
  standard sirens within five years}},
\newblock \bibinfo{journal}{\nat} \bibinfo{volume}{562} (\bibinfo{year}{2018})
  \bibinfo{pages}{545--547}. \DOIprefix\doi{10.1038/s41586-018-0606-0}.
  \href{http://arxiv.org/abs/1712.06531}{{\tt arXiv:1712.06531}}.
\bibitem[{{Sandage}(1962)}]{07-ApJ62-Sandage}
\bibinfo{author}{A.~{Sandage}},
\newblock \bibinfo{title}{{The Change of Redshift and Apparent Luminosity of
  Galaxies due to the Deceleration of Selected Expanding Universes.}},
\newblock \bibinfo{journal}{\apj} \bibinfo{volume}{136} (\bibinfo{year}{1962})
  \bibinfo{pages}{319}. \DOIprefix\doi{10.1086/147385}.
\bibitem[{{Loeb}(1998)}]{08-ApJ98-Loeb}
\bibinfo{author}{A.~{Loeb}},
\newblock \bibinfo{title}{{Direct Measurement of Cosmological Parameters from
  the Cosmic Deceleration of Extragalactic Objects}},
\newblock \bibinfo{journal}{\apj} \bibinfo{volume}{499} (\bibinfo{year}{1998})
  \bibinfo{pages}{L111--L114}. \DOIprefix\doi{10.1086/311375}.
  \href{http://arxiv.org/abs/astro-ph/9802122}{{\tt arXiv:astro-ph/9802122}}.
\bibitem[{{Weltman} et~al.(2020){Weltman}, {Bull}, {Camera}, {Kelley},
  {Padmanabhan}, {Pritchard}, {Raccanelli}, {Riemer-S{\o}rensen}, {Shao},
  {Andrianomena}, {Athanassoula}, {Bacon}, {Barkana}, {Bertone}, {B{\oe}hm},
  {Bonvin}, {Bosma}, {Br{\"u}ggen}, {Burigana}, {Calore}, {Cembranos},
  {Clarkson}, {Connors}, {Cruz-Dombriz}, {Dunsby}, {Fonseca}, {Fornengo},
  {Gaggero}, {Harrison}, {Larena}, {Ma}, {Maartens}, {M{\'e}ndez-Isla},
  {Mohanty}, {Murray}, {Parkinson}, {Pourtsidou}, {Quinn}, {Regis}, {Saha},
  {Sahl{\'e}n}, {Sakellariadou}, {Silk}, {Trombetti}, {Vazza}, {Venumadhav},
  {Vidotto}, {Villaescusa-Navarro}, {Wang}, {Weniger}, {Wolz}, {Zhang}, and
  {Gaensler}}]{09-PASA20-Weltman}
\bibinfo{author}{A.~{Weltman}}, \bibinfo{author}{P.~{Bull}},
  \bibinfo{author}{S.~{Camera}}, \bibinfo{author}{K.~{Kelley}},
  \bibinfo{author}{H.~{Padmanabhan}}, \bibinfo{author}{J.~{Pritchard}},
  \bibinfo{author}{A.~{Raccanelli}}, \bibinfo{author}{S.~{Riemer-S{\o}rensen}},
  \bibinfo{author}{L.~{Shao}}, \bibinfo{author}{S.~{Andrianomena}},
  \bibinfo{author}{E.~{Athanassoula}}, \bibinfo{author}{D.~{Bacon}},
  \bibinfo{author}{R.~{Barkana}}, \bibinfo{author}{G.~{Bertone}},
  \bibinfo{author}{C.~{B{\oe}hm}}, \bibinfo{author}{C.~{Bonvin}},
  \bibinfo{author}{A.~{Bosma}}, \bibinfo{author}{M.~{Br{\"u}ggen}},
  \bibinfo{author}{C.~{Burigana}}, \bibinfo{author}{F.~{Calore}},
  \bibinfo{author}{J.~A.~R. {Cembranos}}, \bibinfo{author}{C.~{Clarkson}},
  \bibinfo{author}{R.~M.~T. {Connors}}, \bibinfo{author}{{\'A}.~d.~l.
  {Cruz-Dombriz}}, \bibinfo{author}{P.~K.~S. {Dunsby}},
  \bibinfo{author}{J.~{Fonseca}}, \bibinfo{author}{N.~{Fornengo}},
  \bibinfo{author}{D.~{Gaggero}}, \bibinfo{author}{I.~{Harrison}},
  \bibinfo{author}{J.~{Larena}}, \bibinfo{author}{Y.~Z. {Ma}},
  \bibinfo{author}{R.~{Maartens}}, \bibinfo{author}{M.~{M{\'e}ndez-Isla}},
  \bibinfo{author}{S.~D. {Mohanty}}, \bibinfo{author}{S.~{Murray}},
  \bibinfo{author}{D.~{Parkinson}}, \bibinfo{author}{A.~{Pourtsidou}},
  \bibinfo{author}{P.~J. {Quinn}}, \bibinfo{author}{M.~{Regis}},
  \bibinfo{author}{P.~{Saha}}, \bibinfo{author}{M.~{Sahl{\'e}n}},
  \bibinfo{author}{M.~{Sakellariadou}}, \bibinfo{author}{J.~{Silk}},
  \bibinfo{author}{T.~{Trombetti}}, \bibinfo{author}{F.~{Vazza}},
  \bibinfo{author}{T.~{Venumadhav}}, \bibinfo{author}{F.~{Vidotto}},
  \bibinfo{author}{F.~{Villaescusa-Navarro}}, \bibinfo{author}{Y.~{Wang}},
  \bibinfo{author}{C.~{Weniger}}, \bibinfo{author}{L.~{Wolz}},
  \bibinfo{author}{F.~{Zhang}}, \bibinfo{author}{B.~M. {Gaensler}},
\newblock \bibinfo{title}{{Fundamental physics with the Square Kilometre
  Array}},
\newblock \bibinfo{journal}{\pasa} \bibinfo{volume}{37} (\bibinfo{year}{2020})
  \bibinfo{pages}{e002}. \DOIprefix\doi{10.1017/pasa.2019.42}.
  \href{http://arxiv.org/abs/1810.02680}{{\tt arXiv:1810.02680}}.
\bibitem[{{Melia}(2016)}]{10-MNRAS16-Melia}
\bibinfo{author}{F.~{Melia}},
\newblock \bibinfo{title}{{Definitive test of the R$_{h}$ = ct universe using
  redshift drift}},
\newblock \bibinfo{journal}{\mnras} \bibinfo{volume}{463}
  (\bibinfo{year}{2016}) \bibinfo{pages}{L61--L63}.
  \DOIprefix\doi{10.1093/mnrasl/slw157}.
  \href{http://arxiv.org/abs/1608.00047}{{\tt arXiv:1608.00047}}.
\bibitem[{{Calabrese} et~al.(2014){Calabrese}, {Martinelli}, {Pandolfi},
  {Cardone}, {Martins}, {Spiro}, and {Vielzeuf}}]{11-PRD14-Cala}
\bibinfo{author}{E.~{Calabrese}}, \bibinfo{author}{M.~{Martinelli}},
  \bibinfo{author}{S.~{Pandolfi}}, \bibinfo{author}{V.~F. {Cardone}},
  \bibinfo{author}{C.~J.~A.~P. {Martins}}, \bibinfo{author}{S.~{Spiro}},
  \bibinfo{author}{P.~E. {Vielzeuf}},
\newblock \bibinfo{title}{{Dark energy coupling with electromagnetism as seen
  from future low-medium redshift probes}},
\newblock \bibinfo{journal}{\prd} \bibinfo{volume}{89} (\bibinfo{year}{2014})
  \bibinfo{pages}{083509}. \DOIprefix\doi{10.1103/PhysRevD.89.083509}.
  \href{http://arxiv.org/abs/1311.5841}{{\tt arXiv:1311.5841}}.
\bibitem[{{Zhang} and {Liu}(2014)}]{12-EPJC14-ZHANG}
\bibinfo{author}{M.-J. {Zhang}}, \bibinfo{author}{W.-B. {Liu}},
\newblock \bibinfo{title}{{Observational constraint on the interacting dark
  energy models including the Sandage-Loeb test}},
\newblock \bibinfo{journal}{European Physical Journal C} \bibinfo{volume}{74}
  (\bibinfo{year}{2014}) \bibinfo{pages}{2863}.
  \DOIprefix\doi{10.1140/epjc/s10052-014-2863-x}.
  \href{http://arxiv.org/abs/1312.0224}{{\tt arXiv:1312.0224}}.
\bibitem[{{Li} et~al.(2013){Li}, {Liao}, {Wu}, {Yu}, and {Zhu}}]{13-PRD13-LI}
\bibinfo{author}{Z.~{Li}}, \bibinfo{author}{K.~{Liao}},
  \bibinfo{author}{P.~{Wu}}, \bibinfo{author}{H.~{Yu}}, \bibinfo{author}{Z.-H.
  {Zhu}},
\newblock \bibinfo{title}{{Probing modified gravity theories with the
  Sandage-Loeb test}},
\newblock \bibinfo{journal}{\prd} \bibinfo{volume}{88} (\bibinfo{year}{2013})
  \bibinfo{pages}{023003}. \DOIprefix\doi{10.1103/PhysRevD.88.023003}.
  \href{http://arxiv.org/abs/1306.5932}{{\tt arXiv:1306.5932}}.
\bibitem[{{Bhattacharjee} and {Sahoo}(2020)}]{14-NewA20-Bhatta}
\bibinfo{author}{S.~{Bhattacharjee}}, \bibinfo{author}{P.~K. {Sahoo}},
\newblock \bibinfo{title}{{Redshift Drift in f(R, T) Gravity}},
\newblock \bibinfo{journal}{\na} \bibinfo{volume}{81} (\bibinfo{year}{2020})
  \bibinfo{pages}{101425}. \DOIprefix\doi{10.1016/j.newast.2020.101425}.
  \href{http://arxiv.org/abs/2005.11163}{{\tt arXiv:2005.11163}}.
\bibitem[{{Martins} et~al.(2016){Martins}, {Martinelli}, {Calabrese}, and
  {Ramos}}]{15-PHD16-Martins}
\bibinfo{author}{C.~J.~A.~P. {Martins}}, \bibinfo{author}{M.~{Martinelli}},
  \bibinfo{author}{E.~{Calabrese}}, \bibinfo{author}{M.~P.~L.~P. {Ramos}},
\newblock \bibinfo{title}{{Real-time cosmography with redshift derivatives}},
\newblock \bibinfo{journal}{\prd} \bibinfo{volume}{94} (\bibinfo{year}{2016})
  \bibinfo{pages}{043001}. \DOIprefix\doi{10.1103/PhysRevD.94.043001}.
  \href{http://arxiv.org/abs/1606.07261}{{\tt arXiv:1606.07261}}.
\bibitem[{{Jimenez} et~al.(2018){Jimenez}, {Raccanelli}, {Verde}, and
  {Matarrese}}]{16-JCAP18-Jimenez}
\bibinfo{author}{R.~{Jimenez}}, \bibinfo{author}{A.~{Raccanelli}},
  \bibinfo{author}{L.~{Verde}}, \bibinfo{author}{S.~{Matarrese}},
\newblock \bibinfo{title}{{Peering beyond the horizon with standard sirens and
  redshift drift}},
\newblock \bibinfo{journal}{\jcap} \bibinfo{volume}{04} (\bibinfo{year}{2018})
  \bibinfo{pages}{002}. \DOIprefix\doi{10.1088/1475-7516/2018/04/002}.
  \href{http://arxiv.org/abs/1711.07984}{{\tt arXiv:1711.07984}}.
\bibitem[{{Yuan} and {Zhang}(2015)}]{17-JCAP15-YUAN}
\bibinfo{author}{S.~{Yuan}}, \bibinfo{author}{T.-J. {Zhang}},
\newblock \bibinfo{title}{{Breaking through the high redshift bottleneck of
  Observational Hubble parameter Data: the Sandage-Loeb signal Scheme}},
\newblock \bibinfo{journal}{\jcap} \bibinfo{volume}{02} (\bibinfo{year}{2015})
  \bibinfo{pages}{025}. \DOIprefix\doi{10.1088/1475-7516/2015/02/025}.
  \href{http://arxiv.org/abs/1311.1583}{{\tt arXiv:1311.1583}}.
\bibitem[{{Koksbang}(2020)}]{53-MNRAS20-Koks}
\bibinfo{author}{S.~M. {Koksbang}},
\newblock \bibinfo{title}{{Observations in statistically homogeneous, locally
  inhomogeneous cosmological toy models without FLRW backgrounds}},
\newblock \bibinfo{journal}{\mnras} \bibinfo{volume}{498}
  (\bibinfo{year}{2020}) \bibinfo{pages}{L135--L139}.
  \DOIprefix\doi{10.1093/mnrasl/slaa146}.
  \href{http://arxiv.org/abs/2008.07108}{{\tt arXiv:2008.07108}}.
\bibitem[{{Koksbang}(2021)}]{54-PRL21-Koks}
\bibinfo{author}{S.~M. {Koksbang}},
\newblock \bibinfo{title}{{Searching for Signals of Inhomogeneity Using
  Multiple Probes of the Cosmic Expansion Rate H (z )}},
\newblock \bibinfo{journal}{\prl} \bibinfo{volume}{126} (\bibinfo{year}{2021})
  \bibinfo{pages}{231101}. \DOIprefix\doi{10.1103/PhysRevLett.126.231101}.
  \href{http://arxiv.org/abs/2105.11880}{{\tt arXiv:2105.11880}}.
\bibitem[{{Heinesen}(2021)}]{19-PRD21-Heinesen}
\bibinfo{author}{A.~{Heinesen}},
\newblock \bibinfo{title}{{Redshift drift as a model independent probe of dark
  energy}},
\newblock \bibinfo{journal}{\prd} \bibinfo{volume}{103} (\bibinfo{year}{2021})
  \bibinfo{pages}{L081302}. \DOIprefix\doi{10.1103/PhysRevD.103.L081302}.
  \href{http://arxiv.org/abs/2102.03774}{{\tt arXiv:2102.03774}}.
\bibitem[{{Piattella} and {Giani}(2017)}]{20-PRD17-Piattella}
\bibinfo{author}{O.~F. {Piattella}}, \bibinfo{author}{L.~{Giani}},
\newblock \bibinfo{title}{{Redshift drift of gravitational lensing}},
\newblock \bibinfo{journal}{\prd} \bibinfo{volume}{95} (\bibinfo{year}{2017})
  \bibinfo{pages}{101301}. \DOIprefix\doi{10.1103/PhysRevD.95.101301}.
  \href{http://arxiv.org/abs/1703.05142}{{\tt arXiv:1703.05142}}.
\bibitem[{{Martins} et~al.(2021){Martins}, {Alves}, {Esteves}, {Lapel}, and
  {Pereira}}]{55-arXiv21-Martins}
\bibinfo{author}{C.~J.~A.~P. {Martins}}, \bibinfo{author}{C.~S. {Alves}},
  \bibinfo{author}{J.~{Esteves}}, \bibinfo{author}{A.~{Lapel}},
  \bibinfo{author}{B.~G. {Pereira}},
\newblock \bibinfo{title}{{Closing the cosmological loop with the redshift
  drift}}  (\bibinfo{year}{2021}) \bibinfo{pages}{arXiv:2110.12242}.
  \href{http://arxiv.org/abs/2110.12242}{{\tt arXiv:2110.12242}}.
\bibitem[{{Lobo} et~al.(2020){Lobo}, {Mimoso}, and {Visser}}]{21-JCAP20-Lobo}
\bibinfo{author}{F.~S.~N. {Lobo}}, \bibinfo{author}{J.~P. {Mimoso}},
  \bibinfo{author}{M.~{Visser}},
\newblock \bibinfo{title}{{Cosmographic analysis of redshift drift}},
\newblock \bibinfo{journal}{\jcap} \bibinfo{volume}{2020}
  (\bibinfo{year}{2020}) \bibinfo{pages}{043}.
  \DOIprefix\doi{10.1088/1475-7516/2020/04/043}.
  \href{http://arxiv.org/abs/2001.11964}{{\tt arXiv:2001.11964}}.
\bibitem[{{Heinesen}(2021)}]{22-PRD21-Heinesen}
\bibinfo{author}{A.~{Heinesen}},
\newblock \bibinfo{title}{{Multipole decomposition of redshift drift:
  Model-independent mapping of the expansion history of the Universe}},
\newblock \bibinfo{journal}{\prd} \bibinfo{volume}{103} (\bibinfo{year}{2021})
  \bibinfo{pages}{023537}. \DOIprefix\doi{10.1103/PhysRevD.103.023537}.
  \href{http://arxiv.org/abs/2011.10048}{{\tt arXiv:2011.10048}}.
\bibitem[{{Corasaniti} et~al.(2007){Corasaniti}, {Huterer}, and
  {Melchiorri}}]{52-PRD07-Corasannti}
\bibinfo{author}{P.-S. {Corasaniti}}, \bibinfo{author}{D.~{Huterer}},
  \bibinfo{author}{A.~{Melchiorri}},
\newblock \bibinfo{title}{{Exploring the dark energy redshift desert with the
  Sandage-Loeb test}},
\newblock \bibinfo{journal}{\prd} \bibinfo{volume}{75} (\bibinfo{year}{2007})
  \bibinfo{pages}{062001}. \DOIprefix\doi{10.1103/PhysRevD.75.062001}.
  \href{http://arxiv.org/abs/astro-ph/0701433}{{\tt arXiv:astro-ph/0701433}}.
\bibitem[{{Liske} et~al.(2008){Liske}, {Grazian}, {Vanzella}, {Dessauges},
  {Viel}, {Pasquini}, {Haehnelt}, {Cristiani}, {Pepe}, {Avila}, {Bonifacio},
  {Bouchy}, {Dekker}, {Delabre}, {D'Odorico}, {D'Odorico}, {Levshakov},
  {Lovis}, {Mayor}, {Molaro}, {Moscardini}, {Murphy}, {Queloz}, {Shaver},
  {Udry}, {Wiklind}, and {Zucker}}]{25-MNRAS08-Liske}
\bibinfo{author}{J.~{Liske}}, \bibinfo{author}{A.~{Grazian}},
  \bibinfo{author}{E.~{Vanzella}}, \bibinfo{author}{M.~{Dessauges}},
  \bibinfo{author}{M.~{Viel}}, \bibinfo{author}{L.~{Pasquini}},
  \bibinfo{author}{M.~{Haehnelt}}, \bibinfo{author}{S.~{Cristiani}},
  \bibinfo{author}{F.~{Pepe}}, \bibinfo{author}{G.~{Avila}},
  \bibinfo{author}{P.~{Bonifacio}}, \bibinfo{author}{F.~{Bouchy}},
  \bibinfo{author}{H.~{Dekker}}, \bibinfo{author}{B.~{Delabre}},
  \bibinfo{author}{S.~{D'Odorico}}, \bibinfo{author}{V.~{D'Odorico}},
  \bibinfo{author}{S.~{Levshakov}}, \bibinfo{author}{C.~{Lovis}},
  \bibinfo{author}{M.~{Mayor}}, \bibinfo{author}{P.~{Molaro}},
  \bibinfo{author}{L.~{Moscardini}}, \bibinfo{author}{M.~T. {Murphy}},
  \bibinfo{author}{D.~{Queloz}}, \bibinfo{author}{P.~{Shaver}},
  \bibinfo{author}{S.~{Udry}}, \bibinfo{author}{T.~{Wiklind}},
  \bibinfo{author}{S.~{Zucker}},
\newblock \bibinfo{title}{{Cosmic dynamics in the era of Extremely Large
  Telescopes}},
\newblock \bibinfo{journal}{\mnras} \bibinfo{volume}{386}
  (\bibinfo{year}{2008}) \bibinfo{pages}{1192--1218}.
  \DOIprefix\doi{10.1111/j.1365-2966.2008.13090.x}.
  \href{http://arxiv.org/abs/0802.1532}{{\tt arXiv:0802.1532}}.
\bibitem[{{Kloeckner} et~al.(2015){Kloeckner}, {Obreschkow}, {Martins},
  {Raccanelli}, {Champion}, {Roy}, {Lobanov}, {Wagner}, and
  {Keller}}]{27-AASKA14-Kloeck}
\bibinfo{author}{H.~R. {Kloeckner}}, \bibinfo{author}{D.~{Obreschkow}},
  \bibinfo{author}{C.~{Martins}}, \bibinfo{author}{A.~{Raccanelli}},
  \bibinfo{author}{D.~{Champion}}, \bibinfo{author}{A.~L. {Roy}},
  \bibinfo{author}{A.~{Lobanov}}, \bibinfo{author}{J.~{Wagner}},
  \bibinfo{author}{R.~{Keller}},
\newblock \bibinfo{title}{{Real time cosmology - A direct measure of the
  expansion rate of the Universe with the SKA}},
\newblock in: \bibinfo{booktitle}{Advancing Astrophysics with the Square
  Kilometre Array (AASKA14)}, \bibinfo{year}{2015}, p.~\bibinfo{pages}{27}.
  \href{http://arxiv.org/abs/1501.03822}{{\tt arXiv:1501.03822}}.
\bibitem[{{Yu} et~al.(2014){Yu}, {Zhang}, and {Pen}}]{28-PRL14-YU}
\bibinfo{author}{H.-R. {Yu}}, \bibinfo{author}{T.-J. {Zhang}},
  \bibinfo{author}{U.-L. {Pen}},
\newblock \bibinfo{title}{{Method for Direct Measurement of Cosmic Acceleration
  by 21-cm Absorption Systems}},
\newblock \bibinfo{journal}{\prl} \bibinfo{volume}{113} (\bibinfo{year}{2014})
  \bibinfo{pages}{041303}. \DOIprefix\doi{10.1103/PhysRevLett.113.041303}.
  \href{http://arxiv.org/abs/1311.2363}{{\tt arXiv:1311.2363}}.
\bibitem[{{Yu} et~al.(2017){Yu}, {Pen}, {Zhang}, {Li}, and
  {Chen}}]{29-RAA17-YU}
\bibinfo{author}{H.-R. {Yu}}, \bibinfo{author}{U.-L. {Pen}},
  \bibinfo{author}{T.-J. {Zhang}}, \bibinfo{author}{D.~{Li}},
  \bibinfo{author}{X.~{Chen}},
\newblock \bibinfo{title}{{Blind search for 21-cm absorption systems using a
  new generation of Chinese radio telescopes}},
\newblock \bibinfo{journal}{\raa} \bibinfo{volume}{17} (\bibinfo{year}{2017})
  \bibinfo{pages}{049}. \DOIprefix\doi{10.1088/1674-4527/17/6/49}.
  \href{http://arxiv.org/abs/1704.04338}{{\tt arXiv:1704.04338}}.
\bibitem[{{Bolejko} et~al.(2019){Bolejko}, {Wang}, and
  {Lewis}}]{23-arXiv19-Bolejko}
\bibinfo{author}{K.~{Bolejko}}, \bibinfo{author}{C.~{Wang}},
  \bibinfo{author}{G.~F. {Lewis}},
\newblock \bibinfo{title}{{Direct detection of the cosmic expansion: the
  redshift drift and the flux drift}}  (\bibinfo{year}{2019})
  \bibinfo{pages}{arXiv:1907.04495}.
  \href{http://arxiv.org/abs/1907.04495}{{\tt arXiv:1907.04495}}.
\bibitem[{{Eikenberry} et~al.(2019){Eikenberry}, {Gonzalez}, {Darling},
  {Liske}, {Slepian}, {Mueller}, {Conklin}, {Fulda}, {Mendes de Oliveira},
  {Bentz}, {Jeram}, {Dong}, {Townsend}, {Izuti Nakazono}, {Quimby}, and
  {Welsh}}]{59-BAAS19-Eiken}
\bibinfo{author}{S.~{Eikenberry}}, \bibinfo{author}{A.~{Gonzalez}},
  \bibinfo{author}{J.~{Darling}}, \bibinfo{author}{J.~{Liske}},
  \bibinfo{author}{Z.~{Slepian}}, \bibinfo{author}{G.~{Mueller}},
  \bibinfo{author}{J.~{Conklin}}, \bibinfo{author}{P.~{Fulda}},
  \bibinfo{author}{C.~{Mendes de Oliveira}}, \bibinfo{author}{M.~{Bentz}},
  \bibinfo{author}{S.~{Jeram}}, \bibinfo{author}{C.~{Dong}},
  \bibinfo{author}{A.~{Townsend}}, \bibinfo{author}{L.~M. {Izuti Nakazono}},
  \bibinfo{author}{R.~{Quimby}}, \bibinfo{author}{W.~{Welsh}},
\newblock \bibinfo{title}{{The Cosmic Accelerometer}},
\newblock in: \bibinfo{booktitle}{Bulletin of the American Astronomical
  Society}, volume~\bibinfo{volume}{51}, \bibinfo{year}{2019}, p.
  \bibinfo{pages}{137}. \href{http://arxiv.org/abs/1907.08271}{{\tt
  arXiv:1907.08271}}.
\bibitem[{{Cooke}(2020)}]{24-MNRAS19-Cooke}
\bibinfo{author}{R.~{Cooke}},
\newblock \bibinfo{title}{{The ACCELERATION programme: I. Cosmology with the
  redshift drift}},
\newblock \bibinfo{journal}{\mnras} \bibinfo{volume}{492}
  (\bibinfo{year}{2020}) \bibinfo{pages}{2044--2057}.
  \DOIprefix\doi{10.1093/mnras/stz3465}.
  \href{http://arxiv.org/abs/1912.04983}{{\tt arXiv:1912.04983}}.
\bibitem[{{Chakrabarti} et~al.(2022){Chakrabarti}, {Gonzalez}, {Eikenberry},
  {Erskine}, {Ishak}, {Kim}, {Linder}, {Nomerotski}, {Pierce}, {Slosar},
  {Stankus}, and {Tsai}}]{60-arXiv22-Chakra}
\bibinfo{author}{S.~{Chakrabarti}}, \bibinfo{author}{A.~H. {Gonzalez}},
  \bibinfo{author}{S.~{Eikenberry}}, \bibinfo{author}{D.~{Erskine}},
  \bibinfo{author}{M.~{Ishak}}, \bibinfo{author}{A.~{Kim}},
  \bibinfo{author}{E.~{Linder}}, \bibinfo{author}{A.~{Nomerotski}},
  \bibinfo{author}{M.~{Pierce}}, \bibinfo{author}{A.~{Slosar}},
  \bibinfo{author}{P.~{Stankus}}, \bibinfo{author}{Y.-D. {Tsai}},
\newblock \bibinfo{title}{{Real-time Cosmology with High Precision Spectroscopy
  and Astrometry}}  (\bibinfo{year}{2022}) \bibinfo{pages}{arXiv:2203.05924}.
  \href{http://arxiv.org/abs/2203.05924}{{\tt arXiv:2203.05924}}.
\bibitem[{{Darling}(2012)}]{26-ApJ12-Darling}
\bibinfo{author}{J.~{Darling}},
\newblock \bibinfo{title}{{Toward a Direct Measurement of the Cosmic
  Acceleration}},
\newblock \bibinfo{journal}{\apj} \bibinfo{volume}{761} (\bibinfo{year}{2012})
  \bibinfo{pages}{L26}. \DOIprefix\doi{10.1088/2041-8205/761/2/L26}.
  \href{http://arxiv.org/abs/1211.4585}{{\tt arXiv:1211.4585}}.
\bibitem[{{Jiao} et~al.(2020){Jiao}, {Zhang}, {Zhang}, {Yu}, {Zhu}, and
  {Li}}]{30-JCAP20-JIAO}
\bibinfo{author}{K.~{Jiao}}, \bibinfo{author}{J.-C. {Zhang}},
  \bibinfo{author}{T.-J. {Zhang}}, \bibinfo{author}{H.-R. {Yu}},
  \bibinfo{author}{M.~{Zhu}}, \bibinfo{author}{D.~{Li}},
\newblock \bibinfo{title}{{Toward a direct measurement of the cosmic
  acceleration: roadmap and forecast on FAST}},
\newblock \bibinfo{journal}{\jcap} \bibinfo{volume}{2020}
  (\bibinfo{year}{2020}) \bibinfo{pages}{054}.
  \DOIprefix\doi{10.1088/1475-7516/2020/01/054}.
  \href{http://arxiv.org/abs/1905.01184}{{\tt arXiv:1905.01184}}.
\bibitem[{{Jimenez} and {Loeb}(2002)}]{61-apj02-Jimen}
\bibinfo{author}{R.~{Jimenez}}, \bibinfo{author}{A.~{Loeb}},
\newblock \bibinfo{title}{{Constraining Cosmological Parameters Based on
  Relative Galaxy Ages}},
\newblock \bibinfo{journal}{\apj} \bibinfo{volume}{573} (\bibinfo{year}{2002})
  \bibinfo{pages}{37--42}. \DOIprefix\doi{10.1086/340549}.
  \href{http://arxiv.org/abs/astro-ph/0106145}{{\tt arXiv:astro-ph/0106145}}.
\bibitem[{{Ruiz-Zapatero} et~al.(2022){Ruiz-Zapatero},
  {Garc{\'\i}a-Garc{\'\i}a}, {Alonso}, {Ferreira}, and
  {Grumitt}}]{62-MN22-RuizZ}
\bibinfo{author}{J.~{Ruiz-Zapatero}},
  \bibinfo{author}{C.~{Garc{\'\i}a-Garc{\'\i}a}},
  \bibinfo{author}{D.~{Alonso}}, \bibinfo{author}{P.~G. {Ferreira}},
  \bibinfo{author}{R.~D.~P. {Grumitt}},
\newblock \bibinfo{title}{{Model-independent constraints on
  {\ensuremath{\Omega}}$_{m}$ and H(z) from the link between geometry and
  growth}},
\newblock \bibinfo{journal}{\mnras} \bibinfo{volume}{512}
  (\bibinfo{year}{2022}) \bibinfo{pages}{1967--1984}.
  \DOIprefix\doi{10.1093/mnras/stac431}.
  \href{http://arxiv.org/abs/2201.07025}{{\tt arXiv:2201.07025}}.
\bibitem[{{Jiao} et~al.(2022){Jiao}, {Borghi}, {Moresco}, and
  {Zhang}}]{63-arx22-Jiao}
\bibinfo{author}{K.~{Jiao}}, \bibinfo{author}{N.~{Borghi}},
  \bibinfo{author}{M.~{Moresco}}, \bibinfo{author}{T.-J. {Zhang}},
\newblock \bibinfo{title}{{A New Observational $H(z)$ Data from Full-Spectrum
  Fitting of Cosmic Chronometers in the LEGA-C Survey}},
\newblock \bibinfo{journal}{arXiv e-prints}  (\bibinfo{year}{2022})
  \bibinfo{pages}{arXiv:2205.05701}.
  \href{http://arxiv.org/abs/2205.05701}{{\tt arXiv:2205.05701}}.
\bibitem[{Alves et~al.(2019)Alves, Leite, Martins, Matos, and
  Silva}]{32-MNRAS19-Alves}
\bibinfo{author}{C.~S. Alves}, \bibinfo{author}{A.~C.~O. Leite},
  \bibinfo{author}{C.~J. A.~P. Martins}, \bibinfo{author}{J.~G.~B. Matos},
  \bibinfo{author}{T.~A. Silva},
\newblock \bibinfo{title}{{Forecasts of redshift drift constraints on
  cosmological parameters}},
\newblock \bibinfo{journal}{\mnras} \bibinfo{volume}{488}
  (\bibinfo{year}{2019}) \bibinfo{pages}{3607--3624}. \URLprefix
  \url{https://doi.org/10.1093/mnras/stz1934}.
  \DOIprefix\doi{10.1093/mnras/stz1934}.
  \href{http://arxiv.org/abs/1907.05151}{{\tt arXiv:1907.05151}}.
\bibitem[{{Meyer} et~al.(2017){Meyer}, {Robotham}, {Obreschkow}, {Westmeier},
  {Duffy}, and {Staveley-Smith}}]{43-PASA17-Meyer}
\bibinfo{author}{M.~{Meyer}}, \bibinfo{author}{A.~{Robotham}},
  \bibinfo{author}{D.~{Obreschkow}}, \bibinfo{author}{T.~{Westmeier}},
  \bibinfo{author}{A.~R. {Duffy}}, \bibinfo{author}{L.~{Staveley-Smith}},
\newblock \bibinfo{title}{{Tracing HI Beyond the Local Universe}},
\newblock \bibinfo{journal}{\pasa} \bibinfo{volume}{34} (\bibinfo{year}{2017})
  \bibinfo{pages}{52}. \DOIprefix\doi{10.1017/pasa.2017.31}.
  \href{http://arxiv.org/abs/1705.04210}{{\tt arXiv:1705.04210}}.
\bibitem[{{Bird} et~al.(2014){Bird}, {Vogelsberger}, {Haehnelt}, {Sijacki},
  {Genel}, {Torrey}, {Springel}, and {Hernquist}}]{33-MNRAS14-Bird}
\bibinfo{author}{S.~{Bird}}, \bibinfo{author}{M.~{Vogelsberger}},
  \bibinfo{author}{M.~{Haehnelt}}, \bibinfo{author}{D.~{Sijacki}},
  \bibinfo{author}{S.~{Genel}}, \bibinfo{author}{P.~{Torrey}},
  \bibinfo{author}{V.~{Springel}}, \bibinfo{author}{L.~{Hernquist}},
\newblock \bibinfo{title}{{Damped Lyman {\ensuremath{\alpha}} absorbers as a
  probe of stellar feedback}},
\newblock \bibinfo{journal}{\mnras} \bibinfo{volume}{445}
  (\bibinfo{year}{2014}) \bibinfo{pages}{2313--2324}.
  \DOIprefix\doi{10.1093/mnras/stu1923}.
  \href{http://arxiv.org/abs/1405.3994}{{\tt arXiv:1405.3994}}.
\bibitem[{{Curran} et~al.(2016){Curran}, {Duchesne}, {Divoli}, and
  {Allison}}]{34-MNRAS16-Curran}
\bibinfo{author}{S.~J. {Curran}}, \bibinfo{author}{S.~W. {Duchesne}},
  \bibinfo{author}{A.~{Divoli}}, \bibinfo{author}{J.~R. {Allison}},
\newblock \bibinfo{title}{{A comparative study of intervening and associated H
  I 21-cm absorption profiles in redshifted galaxies}},
\newblock \bibinfo{journal}{\mnras} \bibinfo{volume}{462}
  (\bibinfo{year}{2016}) \bibinfo{pages}{4197--4207}.
  \DOIprefix\doi{10.1093/mnras/stw1938}.
  \href{http://arxiv.org/abs/1608.01055}{{\tt arXiv:1608.01055}}.
\bibitem[{{Titov} and {Kr{\'a}sn{\'a}}(2018)}]{64-AA18-Titov}
\bibinfo{author}{O.~{Titov}}, \bibinfo{author}{H.~{Kr{\'a}sn{\'a}}},
\newblock \bibinfo{title}{{Measurement of the solar system acceleration using
  the Earth scale factor}},
\newblock \bibinfo{journal}{\aap} \bibinfo{volume}{610} (\bibinfo{year}{2018})
  \bibinfo{pages}{A36}. \DOIprefix\doi{10.1051/0004-6361/201731901}.
  \href{http://arxiv.org/abs/1802.05347}{{\tt arXiv:1802.05347}}.
\bibitem[{{Xu} et~al.(2012){Xu}, {Wang}, and {Zhao}}]{65-AA12-Xu}
\bibinfo{author}{M.~H. {Xu}}, \bibinfo{author}{G.~L. {Wang}},
  \bibinfo{author}{M.~{Zhao}},
\newblock \bibinfo{title}{{The solar acceleration obtained by VLBI
  observations}},
\newblock \bibinfo{journal}{\aap} \bibinfo{volume}{544} (\bibinfo{year}{2012})
  \bibinfo{pages}{A135}. \DOIprefix\doi{10.1051/0004-6361/201219593}.
\bibitem[{{Qian} et~al.(2020){Qian}, {Yao}, {Sun}, {Xu}, {Pan}, and
  {Jiang}}]{66-In20-Qian}
\bibinfo{author}{L.~{Qian}}, \bibinfo{author}{R.~{Yao}},
  \bibinfo{author}{J.~{Sun}}, \bibinfo{author}{J.~{Xu}},
  \bibinfo{author}{Z.~{Pan}}, \bibinfo{author}{P.~{Jiang}},
\newblock \bibinfo{title}{{FAST: Its Scientific Achievements and Prospects}},
\newblock \bibinfo{journal}{The Innovation} \bibinfo{volume}{1}
  (\bibinfo{year}{2020}) \bibinfo{pages}{100053}.
  \DOIprefix\doi{10.1016/j.xinn.2020.100053}.
  \href{http://arxiv.org/abs/2011.13542}{{\tt arXiv:2011.13542}}.
\bibitem[{{Kanekar} and {Chengalur}(2001)}]{36-AA01-Kanekar}
\bibinfo{author}{N.~{Kanekar}}, \bibinfo{author}{J.~N. {Chengalur}},
\newblock \bibinfo{title}{{HI 21 cm absorption in low vec z damped Lyman-alpha
  systems}},
\newblock \bibinfo{journal}{\aap} \bibinfo{volume}{369} (\bibinfo{year}{2001})
  \bibinfo{pages}{42--48}. \DOIprefix\doi{10.1051/0004-6361:20010096}.
  \href{http://arxiv.org/abs/astro-ph/0101402}{{\tt arXiv:astro-ph/0101402}}.
\bibitem[{{Carilli} et~al.(1992){Carilli}, {Perlman}, and
  {Stocke}}]{37-ApJ92-Carilli}
\bibinfo{author}{C.~L. {Carilli}}, \bibinfo{author}{E.~S. {Perlman}},
  \bibinfo{author}{J.~T. {Stocke}},
\newblock \bibinfo{title}{{Discovery of Neutral Hydrogen 21 Centimeter
  Absorption at Redshift 0.25 toward PKS 1413+135}},
\newblock \bibinfo{journal}{\apj} \bibinfo{volume}{400} (\bibinfo{year}{1992})
  \bibinfo{pages}{L13}. \DOIprefix\doi{10.1086/186637}.
\bibitem[{{Lane} et~al.(2000){Lane}, {Briggs}, and {Smette}}]{38-ApJ00-Lane}
\bibinfo{author}{W.~M. {Lane}}, \bibinfo{author}{F.~H. {Briggs}},
  \bibinfo{author}{A.~{Smette}},
\newblock \bibinfo{title}{{Detection of Warm and Cold Phases of the Neutral ISM
  in a Damped Ly{\ensuremath{\alpha}} Absorber}},
\newblock \bibinfo{journal}{\apj} \bibinfo{volume}{532} (\bibinfo{year}{2000})
  \bibinfo{pages}{146--151}. \DOIprefix\doi{10.1086/308578}.
  \href{http://arxiv.org/abs/astro-ph/9911142}{{\tt arXiv:astro-ph/9911142}}.
\bibitem[{{Lane} et~al.(1998){Lane}, {Smette}, {Briggs}, {Rao}, {Turnshek}, and
  {Meylan}}]{39-ApJ98-Lane}
\bibinfo{author}{W.~{Lane}}, \bibinfo{author}{A.~{Smette}},
  \bibinfo{author}{F.~{Briggs}}, \bibinfo{author}{S.~{Rao}},
  \bibinfo{author}{D.~{Turnshek}}, \bibinfo{author}{G.~{Meylan}},
\newblock \bibinfo{title}{{H i 21 Centimeter Absorption in Two Low-Redshift
  Damped Lyalpha Systems}},
\newblock \bibinfo{journal}{\aj} \bibinfo{volume}{116} (\bibinfo{year}{1998})
  \bibinfo{pages}{26--30}. \DOIprefix\doi{10.1086/300422}.
  \href{http://arxiv.org/abs/astro-ph/9803243}{{\tt arXiv:astro-ph/9803243}}.
\bibitem[{{Ger{\'e}b} et~al.(2015){Ger{\'e}b}, {Maccagni}, {Morganti}, and
  {Oosterloo}}]{40-AA15-Gereb}
\bibinfo{author}{K.~{Ger{\'e}b}}, \bibinfo{author}{F.~M. {Maccagni}},
  \bibinfo{author}{R.~{Morganti}}, \bibinfo{author}{T.~A. {Oosterloo}},
\newblock \bibinfo{title}{{The HI absorption ``Zoo''}},
\newblock \bibinfo{journal}{\aap} \bibinfo{volume}{575} (\bibinfo{year}{2015})
  \bibinfo{pages}{A44}. \DOIprefix\doi{10.1051/0004-6361/201424655}.
  \href{http://arxiv.org/abs/1411.0361}{{\tt arXiv:1411.0361}}.
\bibitem[{{Gupta} et~al.(2013){Gupta}, {Srianand}, {Noterdaeme}, {Petitjean},
  and {Muzahid}}]{41-AA13-Gupta}
\bibinfo{author}{N.~{Gupta}}, \bibinfo{author}{R.~{Srianand}},
  \bibinfo{author}{P.~{Noterdaeme}}, \bibinfo{author}{P.~{Petitjean}},
  \bibinfo{author}{S.~{Muzahid}},
\newblock \bibinfo{title}{{21-cm absorption from galaxies at z
  \raisebox{-0.5ex}\textasciitilde 0.3}},
\newblock \bibinfo{journal}{\aap} \bibinfo{volume}{558} (\bibinfo{year}{2013})
  \bibinfo{pages}{A84}. \DOIprefix\doi{10.1051/0004-6361/201321609}.
  \href{http://arxiv.org/abs/1308.4141}{{\tt arXiv:1308.4141}}.
\bibitem[{{Gupta} et~al.(2006){Gupta}, {Salter}, {Saikia}, {Ghosh}, and
  {Jeyakumar}}]{42-MNRAS06-Gupta}
\bibinfo{author}{N.~{Gupta}}, \bibinfo{author}{C.~J. {Salter}},
  \bibinfo{author}{D.~J. {Saikia}}, \bibinfo{author}{T.~{Ghosh}},
  \bibinfo{author}{S.~{Jeyakumar}},
\newblock \bibinfo{title}{{Probing radio source environments via HI and OH
  absorption}},
\newblock \bibinfo{journal}{\mnras} \bibinfo{volume}{373}
  (\bibinfo{year}{2006}) \bibinfo{pages}{972--992}.
  \DOIprefix\doi{10.1111/j.1365-2966.2006.11064.x}.
  \href{http://arxiv.org/abs/astro-ph/0605423}{{\tt arXiv:astro-ph/0605423}}.
\bibitem[{{Becker} et~al.(1995){Becker}, {White}, and
  {Helfand}}]{35-ApJ95-Becker}
\bibinfo{author}{R.~H. {Becker}}, \bibinfo{author}{R.~L. {White}},
  \bibinfo{author}{D.~J. {Helfand}},
\newblock \bibinfo{title}{{The FIRST Survey: Faint Images of the Radio Sky at
  Twenty Centimeters}},
\newblock \bibinfo{journal}{\apj} \bibinfo{volume}{450} (\bibinfo{year}{1995})
  \bibinfo{pages}{559}. \DOIprefix\doi{10.1086/176166}.
\bibitem[{{Fouque} et~al.(1990){Fouque}, {Durand}, {Bottinelli}, {Gouguenheim},
  and {Paturel}}]{44-AAS90-Fouque}
\bibinfo{author}{P.~{Fouque}}, \bibinfo{author}{N.~{Durand}},
  \bibinfo{author}{L.~{Bottinelli}}, \bibinfo{author}{L.~{Gouguenheim}},
  \bibinfo{author}{G.~{Paturel}},
\newblock \bibinfo{title}{{An HI survey of late-type galaxies in the southern
  hemisphere. I. The SGC sample.}},
\newblock \bibinfo{journal}{\aaps} \bibinfo{volume}{86} (\bibinfo{year}{1990})
  \bibinfo{pages}{473}.
\bibitem[{{Koribalski} et~al.(2004){Koribalski}, {Staveley-Smith}, {Kilborn},
  {Ryder}, {Kraan-Korteweg}, {Ryan-Weber}, {Ekers}, {Jerjen}, {Henning},
  {Putman}, {Zwaan}, {de Blok}, {Calabretta}, {Disney}, {Minchin}, {Bhathal},
  {Boyce}, {Drinkwater}, {Freeman}, {Gibson}, {Green}, {Haynes}, {Juraszek},
  {Kesteven}, {Knezek}, {Mader}, {Marquarding}, {Meyer}, {Mould}, {Oosterloo},
  {O'Brien}, {Price}, {Sadler}, {Schr{\"o}der}, {Stewart}, {Stootman}, {Waugh},
  {Warren}, {Webster}, and {Wright}}]{45-AJ04-Kori}
\bibinfo{author}{B.~S. {Koribalski}}, \bibinfo{author}{L.~{Staveley-Smith}},
  \bibinfo{author}{V.~A. {Kilborn}}, \bibinfo{author}{S.~D. {Ryder}},
  \bibinfo{author}{R.~C. {Kraan-Korteweg}}, \bibinfo{author}{E.~V.
  {Ryan-Weber}}, \bibinfo{author}{R.~D. {Ekers}},
  \bibinfo{author}{H.~{Jerjen}}, \bibinfo{author}{P.~A. {Henning}},
  \bibinfo{author}{M.~E. {Putman}}, \bibinfo{author}{M.~A. {Zwaan}},
  \bibinfo{author}{W.~J.~G. {de Blok}}, \bibinfo{author}{M.~R. {Calabretta}},
  \bibinfo{author}{M.~J. {Disney}}, \bibinfo{author}{R.~F. {Minchin}},
  \bibinfo{author}{R.~{Bhathal}}, \bibinfo{author}{P.~J. {Boyce}},
  \bibinfo{author}{M.~J. {Drinkwater}}, \bibinfo{author}{K.~C. {Freeman}},
  \bibinfo{author}{B.~K. {Gibson}}, \bibinfo{author}{A.~J. {Green}},
  \bibinfo{author}{R.~F. {Haynes}}, \bibinfo{author}{S.~{Juraszek}},
  \bibinfo{author}{M.~J. {Kesteven}}, \bibinfo{author}{P.~M. {Knezek}},
  \bibinfo{author}{S.~{Mader}}, \bibinfo{author}{M.~{Marquarding}},
  \bibinfo{author}{M.~{Meyer}}, \bibinfo{author}{J.~R. {Mould}},
  \bibinfo{author}{T.~{Oosterloo}}, \bibinfo{author}{J.~{O'Brien}},
  \bibinfo{author}{R.~M. {Price}}, \bibinfo{author}{E.~M. {Sadler}},
  \bibinfo{author}{A.~{Schr{\"o}der}}, \bibinfo{author}{I.~M. {Stewart}},
  \bibinfo{author}{F.~{Stootman}}, \bibinfo{author}{M.~{Waugh}},
  \bibinfo{author}{B.~E. {Warren}}, \bibinfo{author}{R.~L. {Webster}},
  \bibinfo{author}{A.~E. {Wright}},
\newblock \bibinfo{title}{{The 1000 Brightest HIPASS Galaxies: H I
  Properties}},
\newblock \bibinfo{journal}{\aj} \bibinfo{volume}{128} (\bibinfo{year}{2004})
  \bibinfo{pages}{16--46}. \DOIprefix\doi{10.1086/421744}.
  \href{http://arxiv.org/abs/astro-ph/0404436}{{\tt arXiv:astro-ph/0404436}}.
\bibitem[{{Zhang} et~al.(2021){Zhang}, {Zhu}, {Wu}, {Yu}, {Jiang}, {Yue},
  {Huang}, and {Hao}}]{46-MNRAS21-ZHANG}
\bibinfo{author}{B.~{Zhang}}, \bibinfo{author}{M.~{Zhu}},
  \bibinfo{author}{Z.-Z. {Wu}}, \bibinfo{author}{Q.-Z. {Yu}},
  \bibinfo{author}{P.~{Jiang}}, \bibinfo{author}{Y.-L. {Yue}},
  \bibinfo{author}{M.-L. {Huang}}, \bibinfo{author}{Q.-L. {Hao}},
\newblock \bibinfo{title}{{Extragalactic H I 21-cm absorption line observations
  with the Five-hundred-meter Aperture Spherical radio Telescope}},
\newblock \bibinfo{journal}{\mnras} \bibinfo{volume}{503}
  (\bibinfo{year}{2021}) \bibinfo{pages}{5385--5396}.
  \DOIprefix\doi{10.1093/mnras/stab754}.
  \href{http://arxiv.org/abs/2103.06573}{{\tt arXiv:2103.06573}}.
\bibitem[{{Jiang} et~al.(2019){Jiang}, {Yue}, {Gan}, {Yao}, {Li}, {Pan}, {Sun},
  {Yu}, {Liu}, {Tang}, {Qian}, {Lu}, {Yan}, {Peng}, {Zhang}, {Wang}, {Li}, and
  {Li}}]{47-SCPMA19-JIANG}
\bibinfo{author}{P.~{Jiang}}, \bibinfo{author}{Y.~{Yue}},
  \bibinfo{author}{H.~{Gan}}, \bibinfo{author}{R.~{Yao}},
  \bibinfo{author}{H.~{Li}}, \bibinfo{author}{G.~{Pan}},
  \bibinfo{author}{J.~{Sun}}, \bibinfo{author}{D.~{Yu}},
  \bibinfo{author}{H.~{Liu}}, \bibinfo{author}{N.~{Tang}},
  \bibinfo{author}{L.~{Qian}}, \bibinfo{author}{J.~{Lu}},
  \bibinfo{author}{J.~{Yan}}, \bibinfo{author}{B.~{Peng}},
  \bibinfo{author}{S.~{Zhang}}, \bibinfo{author}{Q.~{Wang}},
  \bibinfo{author}{Q.~{Li}}, \bibinfo{author}{D.~{Li}},
\newblock \bibinfo{title}{{Commissioning progress of the FAST}},
\newblock \bibinfo{journal}{Science China Physics, Mechanics, and Astronomy}
  \bibinfo{volume}{62} (\bibinfo{year}{2019}) \bibinfo{pages}{959502}.
  \DOIprefix\doi{10.1007/s11433-018-9376-1}.
  \href{http://arxiv.org/abs/1903.06324}{{\tt arXiv:1903.06324}}.
\bibitem[{{Baek} et~al.(2015){Baek}, {Park}, {Ahn}, and
  {Choo}}]{48-Analyst15-Beak}
\bibinfo{author}{S.-J. {Baek}}, \bibinfo{author}{A.~{Park}},
  \bibinfo{author}{Y.-J. {Ahn}}, \bibinfo{author}{J.~{Choo}},
\newblock \bibinfo{title}{{Baseline correction using asymmetrically reweighted
  penalized least squares smoothing}},
\newblock \bibinfo{journal}{The Analyst} \bibinfo{volume}{140}
  (\bibinfo{year}{2015}) \bibinfo{pages}{250--257}.
  \DOIprefix\doi{10.1039/C4AN01061B}.
\bibitem[{{Jiang} et~al.(2020){Jiang}, {Tang}, {Hou}, {Liu}, {Kr{\v{c}}o},
  {Qian}, {Sun}, {Ching}, {Liu}, {Duan}, {Yue}, {Gan}, {Yao}, {Li}, {Pan},
  {Yu}, {Liu}, {Li}, {Peng}, {Yan}, and {FAST Collaboration}}]{49-RAA20-JIANG}
\bibinfo{author}{P.~{Jiang}}, \bibinfo{author}{N.-Y. {Tang}},
  \bibinfo{author}{L.-G. {Hou}}, \bibinfo{author}{M.-T. {Liu}},
  \bibinfo{author}{M.~{Kr{\v{c}}o}}, \bibinfo{author}{L.~{Qian}},
  \bibinfo{author}{J.-H. {Sun}}, \bibinfo{author}{T.-C. {Ching}},
  \bibinfo{author}{B.~{Liu}}, \bibinfo{author}{Y.~{Duan}},
  \bibinfo{author}{Y.-L. {Yue}}, \bibinfo{author}{H.-Q. {Gan}},
  \bibinfo{author}{R.~{Yao}}, \bibinfo{author}{H.~{Li}}, \bibinfo{author}{G.-F.
  {Pan}}, \bibinfo{author}{D.-J. {Yu}}, \bibinfo{author}{H.-F. {Liu}},
  \bibinfo{author}{D.~{Li}}, \bibinfo{author}{B.~{Peng}},
  \bibinfo{author}{J.~{Yan}}, \bibinfo{author}{{FAST Collaboration}},
\newblock \bibinfo{title}{{The fundamental performance of FAST with 19-beam
  receiver at L band}},
\newblock \bibinfo{journal}{\raa} \bibinfo{volume}{20} (\bibinfo{year}{2020})
  \bibinfo{pages}{064}. \DOIprefix\doi{10.1088/1674-4527/20/5/64}.
  \href{http://arxiv.org/abs/2002.01786}{{\tt arXiv:2002.01786}}.
\bibitem[{{Czesla} et~al.(2019){Czesla}, {Schr{\"o}ter}, {Schneider}, {Huber},
  {Pfeifer}, {Andreasen}, and {Zechmeister}}]{50-ascl2019-Czesla}
\bibinfo{author}{S.~{Czesla}}, \bibinfo{author}{S.~{Schr{\"o}ter}},
  \bibinfo{author}{C.~P. {Schneider}}, \bibinfo{author}{K.~F. {Huber}},
  \bibinfo{author}{F.~{Pfeifer}}, \bibinfo{author}{D.~T. {Andreasen}},
  \bibinfo{author}{M.~{Zechmeister}}, \bibinfo{title}{{PyA: Python
  astronomy-related packages}}, \bibinfo{year}{2019}.
  \href{http://arxiv.org/abs/1906.010}{{\tt arXiv:1906.010}}.
\bibitem[{{Astropy Collaboration}(2013)}]{51-AA2013-astropy}
\bibinfo{author}{{Astropy Collaboration}},
\newblock \bibinfo{title}{{Astropy: A community Python package for astronomy}},
\newblock \bibinfo{journal}{\aap} \bibinfo{volume}{558} (\bibinfo{year}{2013})
  \bibinfo{pages}{A33}. \DOIprefix\doi{10.1051/0004-6361/201322068}.
  \href{http://arxiv.org/abs/1307.6212}{{\tt arXiv:1307.6212}}.
\bibitem[{{Zhang} et~al.(2020){Zhang}, {Werthimer}, {Zhang}, {Cobb}, {Korpela},
  {Anderson}, {Gajjar}, {Lee}, {Li}, {Pei}, {Zhang}, {Huang}, {Wang}, {Zhu},
  {Duan}, {Zhang}, {Jin}, {Zhu}, and {Li}}]{57-ApJ20-Zhang}
\bibinfo{author}{Z.-S. {Zhang}}, \bibinfo{author}{D.~{Werthimer}},
  \bibinfo{author}{T.-J. {Zhang}}, \bibinfo{author}{J.~{Cobb}},
  \bibinfo{author}{E.~{Korpela}}, \bibinfo{author}{D.~{Anderson}},
  \bibinfo{author}{V.~{Gajjar}}, \bibinfo{author}{R.~{Lee}},
  \bibinfo{author}{S.-Y. {Li}}, \bibinfo{author}{X.~{Pei}},
  \bibinfo{author}{X.-X. {Zhang}}, \bibinfo{author}{S.-J. {Huang}},
  \bibinfo{author}{P.~{Wang}}, \bibinfo{author}{Y.~{Zhu}},
  \bibinfo{author}{R.~{Duan}}, \bibinfo{author}{H.-Y. {Zhang}},
  \bibinfo{author}{C.-j. {Jin}}, \bibinfo{author}{L.-C. {Zhu}},
  \bibinfo{author}{D.~{Li}},
\newblock \bibinfo{title}{{First SETI Observations with China's
  Five-hundred-meter Aperture Spherical Radio Telescope (FAST)}},
\newblock \bibinfo{journal}{\apj} \bibinfo{volume}{891} (\bibinfo{year}{2020})
  \bibinfo{pages}{174}. \DOIprefix\doi{10.3847/1538-4357/ab7376}.
  \href{http://arxiv.org/abs/2002.02130}{{\tt arXiv:2002.02130}}.
\bibitem[{{Kang} et~al.(2022){Kang}, {Zhu}, {Ai}, {Yu}, and
  {Sun}}]{58-arXiv22-Kang}
\bibinfo{author}{J.~{Kang}}, \bibinfo{author}{M.~{Zhu}},
  \bibinfo{author}{M.~{Ai}}, \bibinfo{author}{H.~{Yu}},
  \bibinfo{author}{C.~{Sun}},
\newblock \bibinfo{title}{{Extragalactic HI survey with FAST : First look of
  the pilot survey results}}  (\bibinfo{year}{2022}).
  \href{http://arxiv.org/abs/2204.05840}{{\tt arXiv:2204.05840}}.
\bibitem[{{Allison} et~al.(2022){Allison}, {Sadler}, {Amaral}, {An}, {Curran},
  {Darling}, {Edge}, {Ellison}, {Emig}, {Gaensler}, {Garratt-Smithson},
  {Glowacki}, {Grasha}, {Koribalski}, {Lagos}, {Lah}, {Mahony}, {Mao},
  {Morganti}, {Moss}, {Pettini}, {Pimbblet}, {Power}, {Salas},
  {Staveley-Smith}, {Whiting}, {Wong}, {Yoon}, {Zheng}, and
  {Zwaan}}]{56-PASA22-Allison}
\bibinfo{author}{J.~R. {Allison}}, \bibinfo{author}{E.~M. {Sadler}},
  \bibinfo{author}{A.~D. {Amaral}}, \bibinfo{author}{T.~{An}},
  \bibinfo{author}{S.~J. {Curran}}, \bibinfo{author}{J.~{Darling}},
  \bibinfo{author}{A.~C. {Edge}}, \bibinfo{author}{S.~L. {Ellison}},
  \bibinfo{author}{K.~L. {Emig}}, \bibinfo{author}{B.~M. {Gaensler}},
  \bibinfo{author}{L.~{Garratt-Smithson}}, \bibinfo{author}{M.~{Glowacki}},
  \bibinfo{author}{K.~{Grasha}}, \bibinfo{author}{B.~S. {Koribalski}},
  \bibinfo{author}{C.~d.~P. {Lagos}}, \bibinfo{author}{P.~{Lah}},
  \bibinfo{author}{E.~K. {Mahony}}, \bibinfo{author}{S.~A. {Mao}},
  \bibinfo{author}{R.~{Morganti}}, \bibinfo{author}{V.~A. {Moss}},
  \bibinfo{author}{M.~{Pettini}}, \bibinfo{author}{K.~A. {Pimbblet}},
  \bibinfo{author}{C.~{Power}}, \bibinfo{author}{P.~{Salas}},
  \bibinfo{author}{L.~{Staveley-Smith}}, \bibinfo{author}{M.~T. {Whiting}},
  \bibinfo{author}{O.~I. {Wong}}, \bibinfo{author}{H.~{Yoon}},
  \bibinfo{author}{Z.~{Zheng}}, \bibinfo{author}{M.~A. {Zwaan}},
\newblock \bibinfo{title}{{The First Large Absorption Survey in HI (FLASH): I.
  Science goals and survey design}},
\newblock \bibinfo{journal}{\pasa} \bibinfo{volume}{39} (\bibinfo{year}{2022})
  \bibinfo{pages}{e010}. \DOIprefix\doi{10.1017/pasa.2022.3}.
  \href{http://arxiv.org/abs/2110.00469}{{\tt arXiv:2110.00469}}.

\end{thebibliography}

\end{document}